\RequirePackage{lineno} 

\documentclass[twocolumn,showpacs,preprintnumbers,amsmath,amssymb,nofootinbib]{revtex4}

\usepackage{graphicx}
\usepackage{dcolumn}
\usepackage{bm}
 
\usepackage{epstopdf}

\def\bea{\begin{eqnarray}}
\def\eea{\end{eqnarray}}

\def\pp{\mbox{$p$-$p$}}

\def\pa{\mbox{$p$-A}}
\def\auau{\mbox{Au-Au}}

\def\pbpb{\mbox{Pb-Pb}}
\def\ppb{\mbox{$p$-Pb}}
\def\pn{\mbox{$p$-N}}
\def\aa{\mbox{A-A}}
\def\nn{\mbox{N-N}}

\def\ee{\mbox{$e^+$-$e^-$}}
\def\ppbar{\mbox{$p$-$\bar p$}}

\def\pt{$p_t$}
\def\mt{$m_t$}
\def\yt{$y_t$}

\def\nch{$n_{ch}$}
\def\mmpt{$\bar p_t$}

\begin{document} 

\setlength{\pdfpagewidth}{8.5in}
\setlength{\pdfpageheight}{11in}

\setpagewiselinenumbers
\modulolinenumbers[5]

\preprint{version 2.4}

\title{
Ensemble-mean $\bf p_t$ and hadron production in high-energy nuclear collisions 
}

\author{Thomas A.\ Trainor}\affiliation{CENPA 354290, University of Washington, Seattle, WA 98195}


\date{\today}

\begin{abstract}
A two-component (soft + hard) model (TCM) of hadron production in high-energy nuclear collisions is applied to ensemble-mean $p_t$ (denoted by $\bar p_t$) data for $p$-$p$, $p$-Pb and Pb-Pb collisions from the relativistic heavy ion collider (RHIC) and large hadron collider (LHC). This $\bar p_t$ TCM is directly related to a recently-published TCM for the charge-multiplicity $n_{ch}$ and collision-energy dependence of $p_t$ spectrum data from $p$-$p$ collisions. Multiplicity dependence of the $p$-$p$ spectrum hard component observed in the previous study is consistent with similar behavior for $\bar p_t$ hard component $\bar p_{th}$. $p$-$p$ $\bar p_t$ $n_{ch}$ dependence is observed to follow a {\em noneikonal} trend for the TCM hard component (dijet production $\propto n_{ch}^2$), whereas the trend for Pb-Pb collisions is consistent with the eikonal approximation assumed for the Glauber A-A centrality model. The $p$-Pb trend is intermediate, transitioning from the noneikonal $p$-$p$ trend for more-peripheral collisions to an eikonal trend for more-central collisions. The multiplicity dependence of participant number $N_{part}$ and binary-collision number $N_{bin}$ inferred from  $p$-Pb $\bar p_t$ data differs strongly from a Glauber Monte Carlo model of that system. The rapid increase with $n_{ch}$ and large magnitude of $\bar p_t$ for the $p$-$p$ and $p$-Pb systems suggests that minimum-bias jets (TCM hard component) dominate $\bar p_t$ variation. The trend for $\bar p_{th}$ in Pb-Pb collisions is consistent with quantitative modification of jet formation in more-central A-A collisions.
\end{abstract}


\maketitle

 \section{Introduction}


Measurements of ensemble-mean transverse momentum \pt\ or \mmpt\ for \pp, \pa\ or $d$-A and \aa\ collision systems over a range of collision energies~\cite{ua1mpt,ppprd,alicempt} poses an interesting problem of data interpretation: what mechanism(s) determine hadron production in each of the collision systems? \mmpt\ vs charge multiplicity \nch\ is observed to increase rapidly for \pp\ collisions but much less rapidly for \aa\ collisions. The \pa\ case is intermediate. 

\mmpt\ data from nucleus-nucleus (\aa) collisions have been interpreted conventionally in terms of radial flow (e.g.\ Refs.~\cite{radialflow,alicempt}), nominally a response to large gradients in matter/energy densities~\cite{hydro}. Recent measurements of \mmpt\ vs \nch\ trends in smaller \pa\ or $d$-A systems have revealed much stronger \mmpt\ vs \nch\ dependence~\cite{alicempt}, and those increases have also been interpreted to indicate (possibly stronger) radial flow~\cite{dusling}. Extending that interpretation to the \mmpt\ trend for \pp\ collisions suggests that radial flow may be strongest in that smallest system. 

But progression from large to small systems according to such arguments leads to a paradox within the context of the flow narrative: apparently larger matter/energy gradients must appear in smaller systems, contradicting arguments that have motivated the study of heavy ion collisions~\cite{stoeker}.  And, by inverting the argument if an alternative (e.g.\ jet-related) mechanism dominates the \mmpt\ trend in \pp\ collisions that same mechanism could prevail in \aa\ collisions, misinterpreted there as radial flow~\cite{nohydro}.

Inference of underlying hadron production mechanisms from their manifestations in yields, spectra and correlations is a central goal for analysis of high-energy nuclear collisions. Candidate mechanisms range from projectile-nucleon dissociation and scattered-parton fragmentation to dijets~\cite{kn,hijing,ppprd,fragevo,pythia} to multiple parton interactions (MPI)~\cite{mpitheory,mpiexp} to strong rescattering of partons and hadrons within a dense medium~\cite{ampt} leading to hydrodynamic flows~\cite{hydro,bw,epos}. Emergence of  novel effects in recent high-energy data for smaller \pa\ or $d$-A collision systems from the relativistic heavy ion collider (RHIC) and large hadron collider (LHC) has stimulated debate. Comparisons of \mmpt\ data with Monte Carlo models have been inconclusive~\cite{alicempt}. More-differential data analysis and an improved data model are clearly needed (see Sec.~\ref{choices}).
  
%


The  two-component (soft + hard) model (TCM) of hadron production near midrapidity in high energy nuclear collisions provides a simple framework for analysis and interpretation, with components well defined both mathematically and physically~\cite{ppprd,anomalous,ppquad}. 
In a previous TCM study of $\bar p_t$ trends~\cite{alicetommpt} data from three collision systems at LHC energies~\cite{alicempt} were described by a TCM based on assumptions imposed by limited information. The study assumed that parameter $\alpha$ relating TCM soft and hard components is independent of collision energy and that TCM $\bar p_t$ hard component $\bar p_{th}$ is approximately independent of \nch\ at a given energy. The study established that the TCM can provide a precise description of \mmpt\ data and suggested that the dominant mechanism for \mmpt\ variation at midrapidity is minimum-bias (MB) dijets. A recent study of \pp\ \pt\ spectrum evolution with \nch\ and collision energy~\cite{alicetomspec} has provided additional information on the energy dependence of parameter $\alpha$ and \nch\ dependence of the TCM hard component, among other details, that motivated the updated \mmpt\ study reported here.


The main focus of the present study is \mmpt\ as an integral measure of \pt\ spectrum structure.  This study  reexamines recent \mmpt\ vs \nch\ data from the LHC for \pp, \ppb\ and \pbpb\ collisions~\cite{alicempt}, their variation with collision system A-B, collision energy $\sqrt{s_{NN}}$ from 200 GeV to 7 TeV and charged-hadron multiplicity \nch. The study  confirms that dijet production in \pp\ collisions has a quadratic relation to \nch\ soft component $n_s$. The \pp\ \mmpt\  TCM hard component is directly related to properties of independently-measured jet spectra and jet fragmentation which confirms the role of MB dijets in \pp\ \mmpt\ data. The \ppb\ \mmpt\ TCM trend is simply explained as transitioning from \pp\ to \aa\ behavior as the effective number of nucleon participant pairs becomes significantly greater than one. The \pbpb\ \mmpt\ TCM is a limiting case of the \ppb\ model wherein a Glauber description of  \aa\ geometry dominates for more-central collisions but not for peripheral collisions. \mmpt\ systematics in \aa\ collisions, specifically for the \mmpt\ hard component, reflect  modification of the jet contribution to \pt\ spectra for more-central collisions as expected. The TCM for \mmpt\ data suggests that for any collision system the variation of \mmpt\ with \nch\ or centrality is dominated by MB dijet production.

This article is arranged as follows:
Sec.~\ref{2comp} introduces the general TCM for hadron production near midrapidity in high-energy nuclear collisions.
Sec.~\ref{pptcmm} describes the TCM for \pt\ spectra and \mmpt\ data from \pp\ collisions.
Sec.~\ref{ppb} derives a TCM for \mmpt\ data from \ppb\ collisions.
Sec.~\ref{pbpb} presents similar results for \pbpb\ collisions.
Sec.~\ref{syserr} discusses systematic uncertainties.
Secs.~\ref{disc} and~\ref{summ} present discussion and summary, and App.~\ref{equations} describes a parametrization of \nch\ dependence of the \pp\ spectrum hard component and corresponding \mmpt\ trend.

\section{The two-component model} \label{2comp}

The TCM as utilized in this study emerged from inductive analysis of \pp\ \pt\ spectra. Its two components correspond to basic hadron production mechanisms emerging from several decades of high-energy physics experiments. The TCM provides a simple and accurate description of a broad array of collision data. In this section the basic model is introduced in the context of \pp\ collisions.

\subsection{Choices among analysis methods} \label{choices}

Analysis of collision data assumes a direct connection between underlying mechanisms and the structure of the hadronic final state in  the form of yields, spectra, correlations, and especially jets as hadron correlations. However, how to characterize final-state structures statistically and how to interpret results in terms of physical mechanisms remains an open question. Competing analysis methods may lead to disparate conclusions~\cite{mbdijets}.

The hadronic final state is characterized by various measures including \pt\ spectra and statistical quantities derived therefrom. Whereas differential \pt\ spectra carry more information, integral statistical measures  such as ensemble-mean \mmpt\ are capable of greater precision due to integration. Differential spectrum structure may suggest likely mechanisms whereas integral measures may test hypotheses more precisely. Total \pt\ ($P_t$) and hadron charge (\nch) integrated within some acceptance are {\em extensive} measures whereas ratio $\bar p_t = \bar P_t / n_{ch}$ is an {\em intensive} measure that may conceal important underlying trends  [see Eq.~(\ref{ppmpttcm}) below]. While methods imposing special conditions on data (e.g.\ a ``trigger'' \pt\ cut) are sometimes preferred, minimum-bias analysis (no special conditions) is essential to explore a larger context~\cite{mbdijets}. 

Optimum analysis methods may be suggested by data trends. For instance, the TCM for hadron production near midrapidity in high energy nuclear collisions used in the present study was derived inductively via phenomenological analysis of yield~\cite{jetspec}, spectrum~\cite{ppprd,hardspec} and correlation~\cite{anomalous,ppquad} data from high-energy collisions. The TCM represents the observation that hadron production proceeds by two mechanisms: (a) projectile-nucleon dissociation to charge-neutral hadron pairs (soft) and (b) scattered-parton fragmentation to correlated hadron jets (hard).  A TCM reference may be defined in terms of {\em linear superposition} of a fundamental process: low-$x$ parton-parton interactions within \pp\ collisions or nucleon-nucleon (\nn) interactions within \pa\ and \aa\ collisions. Deviations from a TCM reference may then reveal novelty (e.g.\ nonlinearity) in a composite system. 



\subsection{TCM description of p-p $\bf p_t$ or $\bf y_t$ spectra} \label{aa1}

The TCM for \pp\ collision data emerged from analysis of 200 GeV \pt\ spectrum data. Systematic analysis of the \nch\ dependence of \pt\ spectra from 200 GeV \pp\ collisions described in Ref.~\cite{ppprd} led to a compact phenomenological TCM with {\em approximate factorization} of multiplicity \nch\ and {transverse-rapidity} \yt\ dependence in the form
\bea \label{a1x}
 \frac{d^2n_{ch}}{y_t dy_t d\eta} \approx \bar \rho_0(y_t,n_{ch})&=& S_{pp}(y_t,n_{ch}) + H_{pp}(y_t,n_{ch})
\\ \nonumber
&\approx& \bar \rho_s(n_{ch}) \hat  S_0(y_t) + \bar \rho_h(n_{ch}) \hat H_0(y_t)
\eea
with mean angular densities $\bar \rho_x = n_x / \Delta \eta$. Transverse rapidity $y_t \equiv \ln[(p_t + m_t)/m_h]$ with transverse mass defined by $m_t^2 = p_t^2 + m_h^2$ provides improved visual access to spectrum structure at lower \pt\ or \yt\ (for unidentified hadrons pion mass $m_h = m_\pi$ is assumed). Unit-integral soft-component model $\hat S_0(m_t)$ is consistent with a L\'evy distribution on \mt\ with exponent  $n_0 \approx 12.5$ and slope parameter $T_0 \approx 145$ MeV, while peaked hard-component model  $\hat H_0(y_t)$ is well approximated by a Gaussian on \yt\ centered near $y_t \approx 2.65$ ($p_t \approx 1$ GeV/c) with exponential (power-law) tail reflecting an underlying jet energy spectrum~\cite{fragevo,jetspec2}. Conversion from \pt\ or \mt\ to \yt\ (or the reverse) is accomplished with Jacobian factor $p_t m_t/  y_t$. $\hat S_0(y_t)$ is nearly constant at lower \yt\ making extrapolation to zero especially simple on that variable.

\begin{figure}[h]
\includegraphics[width=1.65in,height=1.65in]{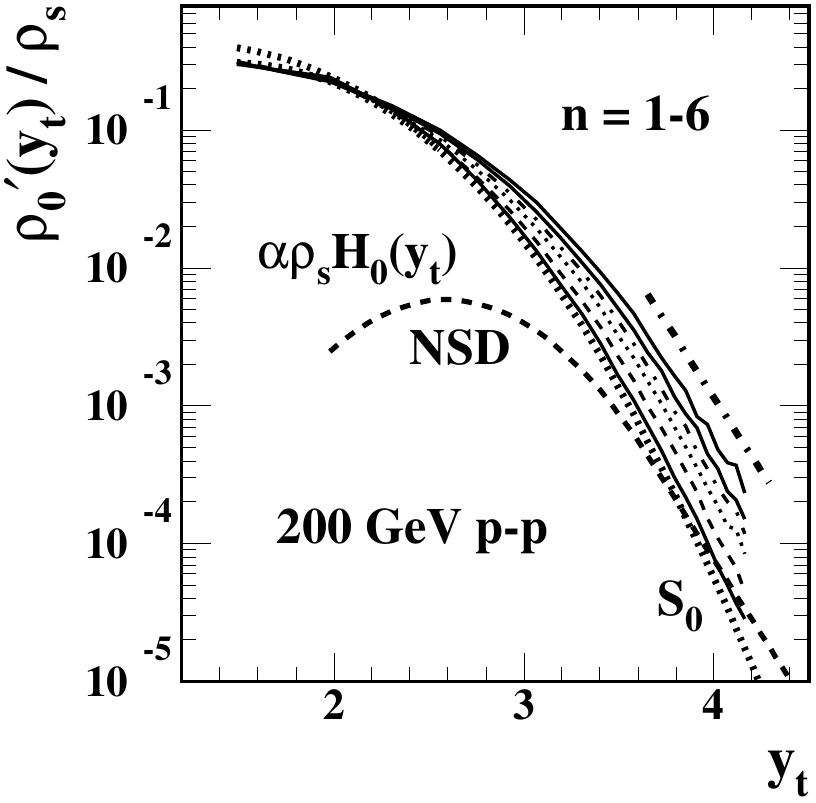}
\includegraphics[width=1.68in]{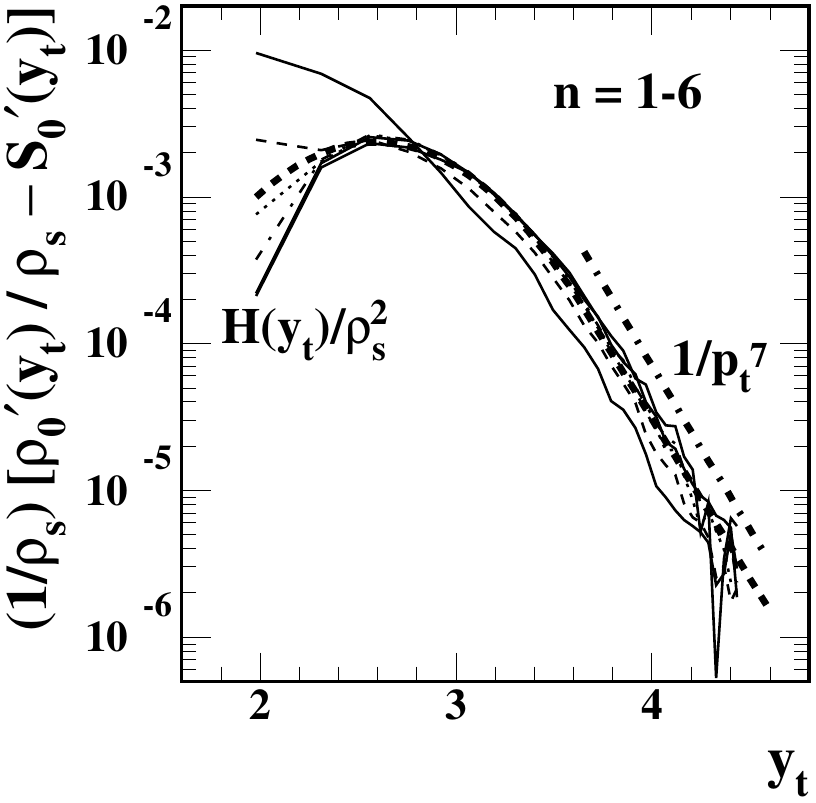}
\caption{\label{pp1}
Left: \yt\ spectra for six charge-multiplicity classes of 200 GeV \pp\ collisions (thin curves) compared to fixed reference $\hat S_0(y_t)$ (bold dotted)~\cite{ppquad}. The bold dashed curve is corresponding hard-component model $\alpha\bar \rho_s \hat H_0(y_t)$ for NSD \pp\ collisions.
Right: Hard-component distributions derived from spectra at left in the form $H(n_{ch},y_t)/\bar \rho_s^2$ (several line styles) compared to fixed reference $\alpha \hat H_0(y_t)$ (bold dashed).  The dash-dotted line in each panel indicates a power-law trend $\approx 1/p_t^7$ corresponding to the underlying jet energy spectrum.
}  
\end{figure}

Figure~\ref{pp1} (left) shows \yt\ spectra for six \pp\ multiplicity classes integrated over acceptance $\Delta \eta = 2$ and $2\pi$ and normalized by soft-component density $\bar \rho_s$. Empirically, all spectra are observed to coincide at lower \yt\ if normalized by $\bar \rho_s \approx \bar \rho_0 - \alpha \bar \rho_0^2$ for some $\alpha \approx O(0.01)$~\cite{ppprd}. The upper bold dotted curve is the same soft-component L\'evy function $\hat  S_0(y_t)$ noted above as a TCM reference.

Figure~\ref{pp1} (right) shows hard-component data inferred from the left panel via  Eq.~(\ref{a1x})  in the form $H_{pp}(y_t,n_{ch}) / \bar \rho_s^2$ (thin curves) compared with a fixed Gaussian model function in the form $\alpha \hat H_0(y_t)$ (bold dashed) with centroid $\bar y_t \approx 2.65$ and width $\sigma_{y_t} \approx 0.45$ and with coefficient $\alpha \approx 0.006$ determined by the data-model comparison. 

Integration of Eq.~(\ref{a1x}) over \yt\ and angular average over some acceptance $\Delta \eta$ near midrapidity results in the angular density TCM $\bar \rho_0 = \bar \rho_s + \bar \rho_h$. Spectrum~\cite{ppprd} and angular-correlation~\cite{ppquad} data reveal that soft and hard angular densities are related by $\bar \rho_h = \alpha \bar \rho_s^2$ with $\alpha \approx 0.006$ within $\Delta \eta = 2$ at 200 GeV. The two relations are equivalent to a quadratic equation that uniquely defines $\bar \rho_s$ and $\bar \rho_h$ in terms of $\bar \rho_0$ (when corrected for inefficiencies).

The soft-component L\'evy distribution $\hat S_0(m_t)$ is similar in form to a ``power-law''  model function that has been used  to fit \pt\ or \mt\ spectra~\cite{ua1mpt}. However, a single L\'evy function alone cannot describe \pp\ spectra accurately, as demonstrated in Ref.~\cite{ppprd}. The two-component spectrum model of Eq.~(\ref{a1x}) is {\em necessary} to describe an ensemble of $p_t$ spectra spanning a significant multiplicity interval.

\subsection{TCM description of two-particle correlations}

Physical interpretation of two phenomenological TCM components (isolated via \nch\ dependence) has proceeded by several routes: 
(a) correspondence with predictions based on measurements of isolated-jet properties~\cite{fragevo},
(b) correspondence with other experimental results (e.g.\ fluctuation measurements~\cite{ptscale}) and
(c) correspondence with MB two-particle correlations on $(y_{t},y_{t})$~\cite{porter2,porter3} and $(\eta_\Delta,\phi_\Delta)$~\cite{ppquad,anomalous}.
Two-particle correlations on transverse-rapidity space $(y_{t},y_{t})$ or angle-difference space $(\eta_\Delta,\phi_\Delta)$ [difference variables on pseudorapidity and azimuth $(\eta,\phi)$] provide direct evidence for a two-component description of hadron production.

\begin{figure}[h]
  \includegraphics[width=1.65in,height=1.4in]{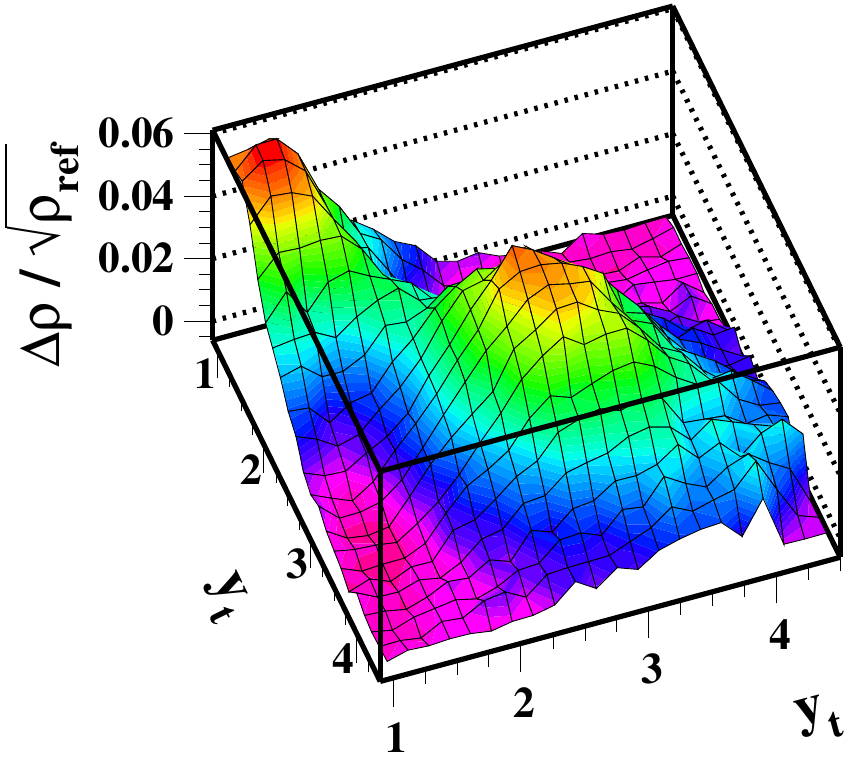}
 \includegraphics[width=1.65in,height=1.4in]{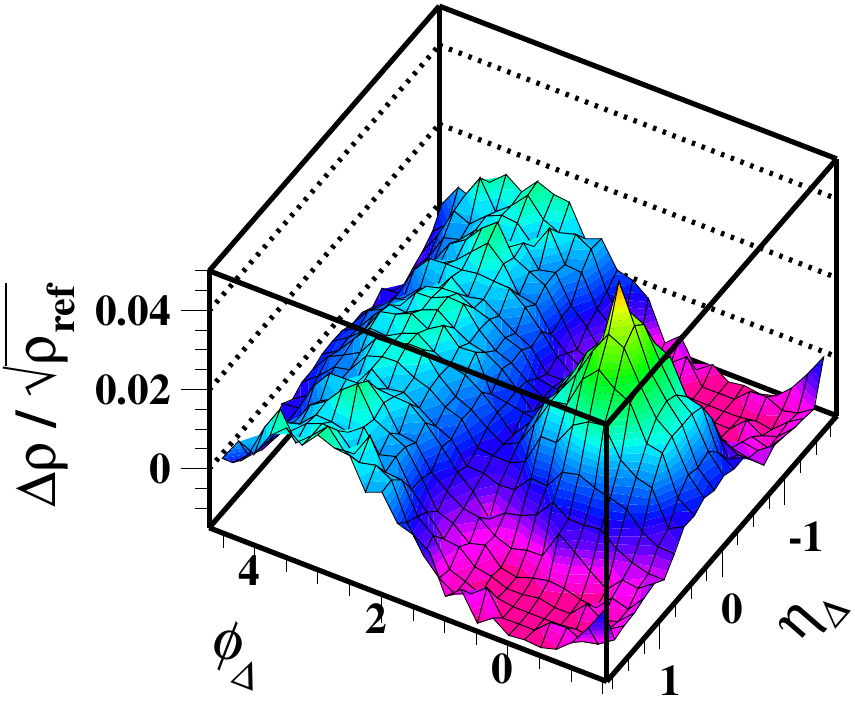}
 \caption{\label{ppcorr} (Color online)  Two-particle correlations on $(y_t, y_t)$ and $(\eta,\phi)$~\cite{porter2,porter3}.
Left: Minimum-bias correlated-pair density on 2D transverse-rapidity space $(y_{t},y_{t})$ from 200 GeV \pp\ collisions showing soft (smaller \yt) and hard (larger \yt) components as peak structures.
Right:  Correlated-pair density on 2D angular difference space $(\eta_\Delta,\phi_\Delta)$. Although hadrons are selected with $p_t \approx 0.6$ GeV/c ($y_t \approx 2$) features expected for dijets are still observed: (i) same-side 2D peak representing intrajet correlations and (ii) away-side 1D peak on azimuth representing interjet (back-to-back jet) correlations. 
 }   
 \end{figure}

Figure~\ref{ppcorr} (left) shows correlations on $(y_{t},y_{t})$ from 200 GeV NSD \pp\ collisions for $p_t \in [0.15,6]$ GeV/c ($y_t \in [1,4.5]$)~\cite{porter2,porter3}. Two peaked features can be identified as TCM soft and hard components as follows. The lower-\yt\ peak falls mainly below 0.5 GeV/c ($y_t < 2$) and consists exclusively of unlike-sign (US) pairs. Corresponding angular correlations consist of a narrow 1D peak on $\eta_\Delta$ centered at the origin. The combination suggests longitudinal fragmentation of low-$x$ gluons to {\em charge-neutral} hadron pairs closely spaced on $\eta$ and consistent with spectrum soft component $S_{pp}(y_t,n_{ch})$ in Eq.~(\ref{a1x}). The higher-\yt\ peak extends mainly above 0.5 GeV/c with mode near $p_t = 1$ GeV/c ($y_t \approx 2.65$) and is consistent with spectrum hard component $H_{pp}(y_t,n_{ch})$ in Eq.~(\ref{a1x}).

Figure~\ref{ppcorr} (right) shows angular correlations for the same collision system with the condition $p_t \approx 0.6$ GeV/c, i.e.\ near the lower boundary of the $(y_t,y_t)$ hard component in the left panel. Despite the low hadron momentum the observed angular correlations exhibit structure expected for jets: a same-side (SS, $|\phi_\Delta| < \pi/2$) 2D peak representing {\em intra}\,jet  correlations and an away-side (AS, $|\phi_\Delta - \pi| < \pi/2$) 1D peak representing  {\em inter}\,jet (back-to-back jet) correlations. The SS peak is dominated by US pairs while the AS peak shows no preference, consistent with fragmentation of back-to-back {\em charge-neutral} gluons.
 In this study the TCM for \pp\ collisions (with extensions to \ppb\ and \pbpb) is applied to LHC \mmpt\ data.

\section{$\bf \bar p_t$ TCM for $\bf p$-$\bf p$ collisions}  \label{pptcmm}

As noted above, Ref.~\cite{alicetommpt} established that a TCM for ratio $\langle p_t \rangle \rightarrow \bar p_t = \bar P_t / n_{ch}$ provides a good description of LHC data from \pp\ and \pbpb\ collisions at several energies, and the TCM provides hints as to the nature of \ppb\ \mmpt\ variation -- transitioning from a \pp\ trend to a \pbpb\ trend with increasing \nch. 

However, certain assumptions for the analysis in Ref.~\cite{alicetommpt} arose from lack of information about spectrum structure. The TCM parameter $\alpha$ relating soft and hard densities $\bar \rho_s$ and $\bar \rho_h$ was well-established for 200 GeV \pp\ collisions~\cite{ppprd,ppquad} but undefined for other energies and was therefore assumed to be independent of energy. Although the ensemble-mean hard component $\bar p_{th}$ for the 200 GeV spectrum was known the values for other energies were not available to constrain the previous \mmpt\ analysis.

Reference~\cite{alicetomspec} reports a recent study of \pp\ \pt\ spectra for a range of energies from 17 GeV to 13 TeV wherein a detailed TCM is developed for spectrum variation with \nch\ and $\sqrt{s}$. Soft component $\hat S_0(m_t,\sqrt{s})$ varies only weakly with energy and not at all with \nch, but hard component $\hat H_0(y_t,n_s,\sqrt{s})$ varies strongly with energy (consistent with jet properties) and significantly with \nch\ (as established with 200 GeV and 13 TeV spectra). 
\pt\ spectrum evolution must have a direct correspondence with \mmpt\ trends. Results from Ref.~\cite{alicetomspec} reviewed next are incorporated in this revised \mmpt\ analysis updating Ref.~\cite{alicetommpt}.

\subsection{Recent results from $\bf p$-$\bf p$ $\bf p_t$ spectrum studies}

The dijet production trend $\bar \rho_h \propto \bar \rho_s^2$ inferred from \pp\ hadron spectra  combined with $\bar \rho_s \propto \ln(\sqrt{s}/\text{10 GeV})$~\cite{alicetomspec} describe jet-spectrum energy trends over large \pp\ collision-energy and jet-energy intervals~\cite{jetspec2}. Predicted jet-spectrum trends can then be combined with measured and parametrized fragmentation functions (FFs) to predict  spectrum hard component $H(p_t,n_{ch},\sqrt{s})$~\cite{fragevo}. 

\begin{figure}[h]
  \includegraphics[width=1.66in]{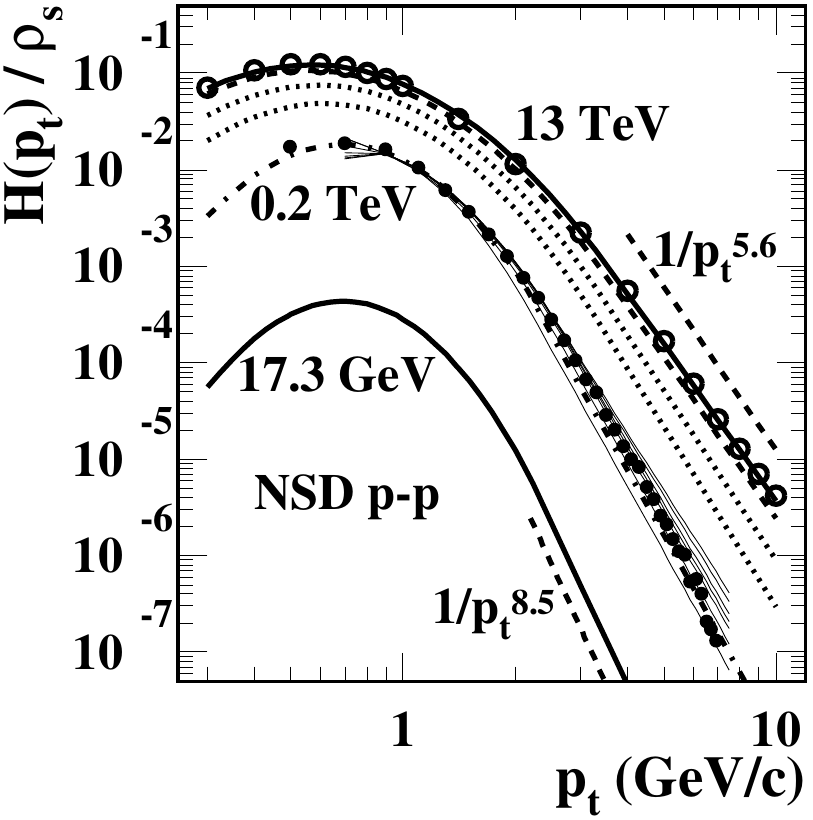}
   \includegraphics[width=1.64in]{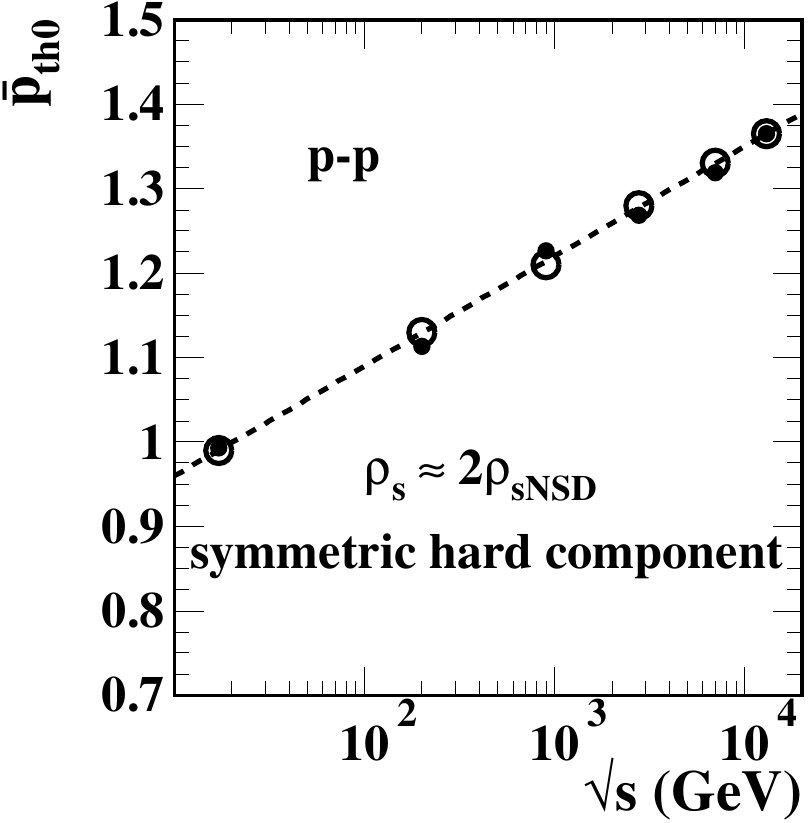}
\caption{\label{enrat3}
Left: A survey of spectrum hard components over the currently accessible energy range from threshold of dijet production (10 GeV) to LHC top energy (13 TeV). The curves are determined by TCM parameters for NSD \pp\ collisions from Ref.~\cite{alicespec}. The 200 GeV fine solid curves illustrate \nch\ dependence. The points are from Refs.~\cite{ppquad} (200 GeV) and \cite{alicespec} (13 TeV).
Right: Values for $\bar p_{th0}$ (solid points) inferred from a symmetric TCM hard-component model with parameters corresponding to $\bar \rho_s \approx 2 \bar \rho_{sNSD}$. The dashed line is a fit by eye and open circles are values used in the present study.
 }  
\end{figure}

Figure~\ref{enrat3} (left) shows ratio $H(p_t,\sqrt{s}) / \bar \rho_s(\sqrt{s}) \approx \alpha(\sqrt{s}) \bar \rho_s(\sqrt{s}) \hat H_0(p_t,\sqrt{s})$ for NSD \pp\ collisions measuring the spectrum hard component  {\em per soft-component hadron} corresponding (by hypothesis) to dijet production via participant low-$x$ gluons. The 13 TeV TCM solid curve is compared to spectrum data (open points). The two dotted curves are for 0.9 and 2.76 TeV and the dashed curve is for 7 TeV. The 200 GeV summary includes \nch\ dependence of $\hat H_0(y_t,n_{ch})$ for seven multiplicity classes (thin solid curves). Corresponding data (solid points) represent NSD \pp\ collisions. Isolated hard components clarify spectrum energy evolution and its relation to dijet production. The predictions for six collision energies (curves) are compared to data from four energies (13, 7, 0.9 and 0.2 TeV) in Ref.~\cite{jetspec2}. The overall result is a comprehensive description of dijet contributions to \pt\ spectra vs \pp\ collision energy over three orders of magnitude.

Figure~\ref{enrat3} (right) shows ensemble-mean $\bar p_{th} = \int dp_t p_t^2 H(p_t,\sqrt{s}) /  \int dp_t p_t H(p_t,\sqrt{s})$ for six energies (solid points) derived from a TCM hard-component model based on a {\em symmetric} Gaussian (see Sec.~\ref{nchvar}) and corresponding to model parameters for $\bar \rho_s \approx 2 \bar \rho_{sNSD}$ from Ref.~\cite{alicetomspec} denoted by $\bar p_{th} \rightarrow \bar p_{th0}$. The dashed line is a fit to the solid points with $\log(\sqrt{s})$ dependence. The open points are $\bar p_{th0}$ values interpolated from the dashed line for use in the present study.

Parameter $\alpha$ that relates soft and hard components of hadron yields in \pp\ collisions is defined by $\bar \rho_h \approx \alpha \bar \rho_s^2$~\cite{ppprd}. It is also related to jet systematics by
\bea
 \alpha \bar \rho_{sNSD}^2 \approx \bar \rho_{hNSD} &\approx& \epsilon(\Delta \eta) f_{NSD}  2\bar n_{ch,j},
\eea
where $2\bar n_{ch,j}$ is the mean hadron fragment multiplicity per dijet averaged over a jet spectrum for given collision energy~\cite{eeprd} and $f_{NSD} = (1/\sigma_{NSD}) d\sigma_{jet} /  d\eta$ is the dijet frequency and $\eta$ density per NSD \pp\ collision~\cite{fragevo}. The energy trends for those quantities, inferred from isolated-jet data, can be used to predict an energy trend for $\alpha$. 



Combining various elements defined below the predicted $\alpha(\sqrt{s})$ trend is
\bea \label{alphas}
\alpha(\sqrt{s}) &\approx& \frac{\epsilon(\Delta \eta)}{\sigma_{NSD}} \frac{d\sigma_{jet}}{d\eta} \frac{2\bar n_{ch,j}}{\bar \rho_{sNSD}^2}
\\ \nonumber
\\ \nonumber
 &\approx& \frac{0.03 \Delta y_{max}}{32 + \Delta y_b^2} \times 2\bar n_{ch,j}(\sqrt{s}),
\eea
where $\Delta y_b = \ln(\sqrt{s}/\text{10 GeV})$ represents an observed cutoff of dijet production near $\sqrt{s} = 10$ GeV, and  $\Delta y_{max} = \ln(\sqrt{s}/\text{6 GeV})$ responds to an inferred infrared cutoff of jet energy spectra near $E_{jet} = 3$ GeV~\cite{jetspec2}.  Dijet acceptance factor $\epsilon \approx 0.6$ is an estimate for  $\Delta \eta = 1.5$ - 2~\cite{jetspec} (but approaches 0.5 for smaller $\Delta \eta$), and $\bar \rho_{sNSD} \approx 0.8 \Delta y_b$~\cite{alicespec}.  Systematic variation of $\bar n_{ch,j}$ with \pp\ collision energy is not well-defined by data at lower energies. In Ref.~\cite{alicespec} a simple $\ln(\sqrt{s})$ trend in the form $2\bar n_{ch,j}(\sqrt{s}) \approx 3(1+\Delta y_{max}/10)$ is consistent with data.

\begin{figure}[h]
  \includegraphics[width=1.60in]{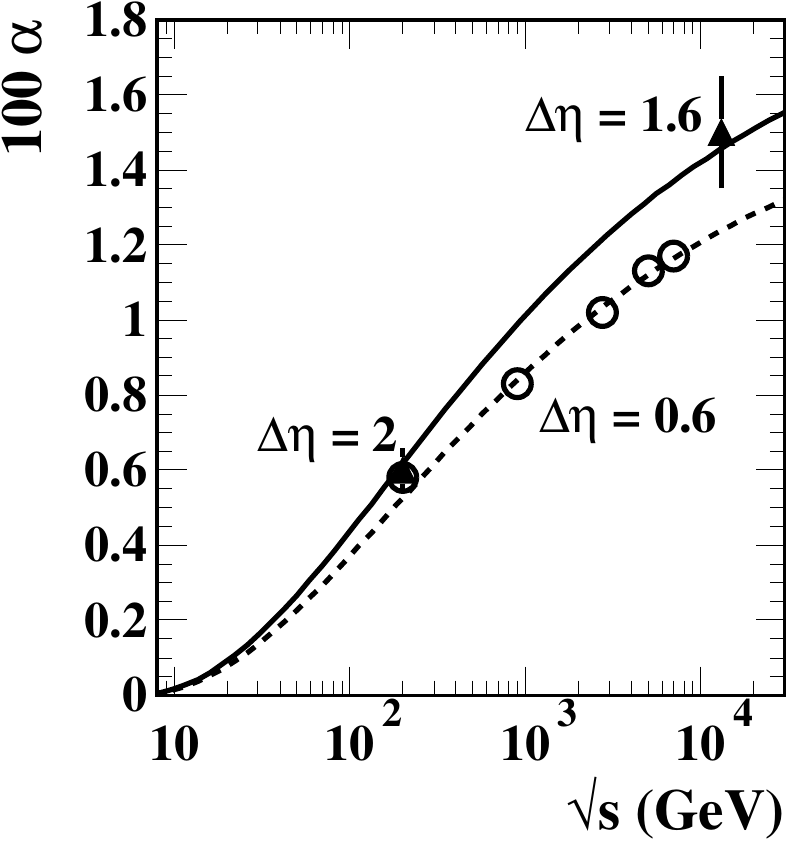}
  \includegraphics[width=1.69in]{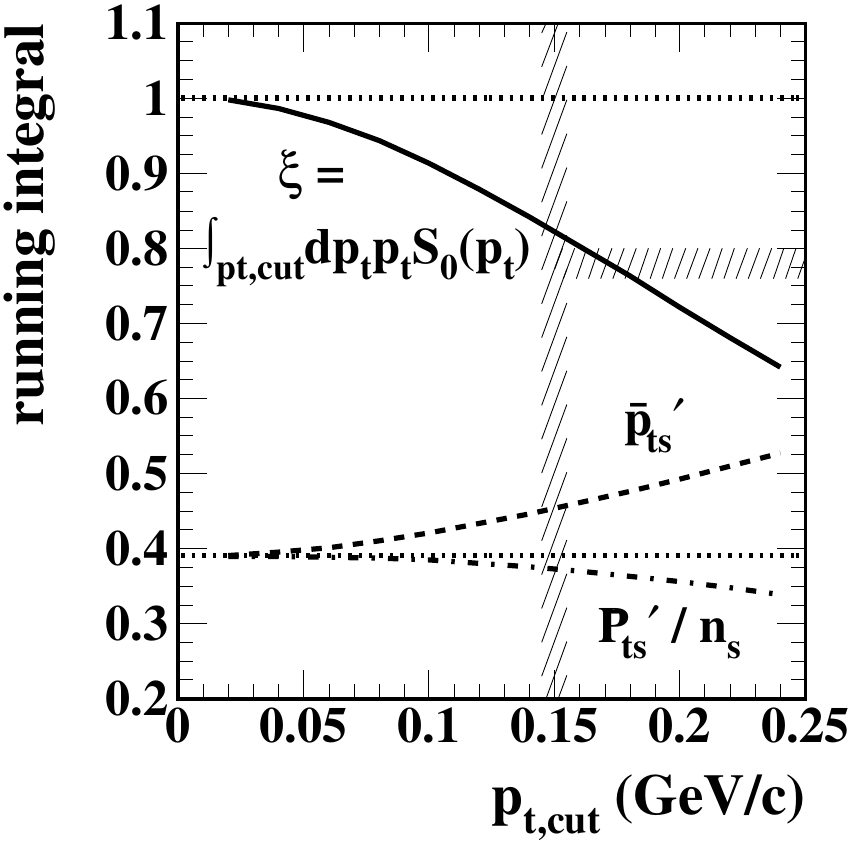}
\caption{\label{params}
Left: TCM hard-soft ratio parameter $\alpha$ determined by analysis of spectrum ratio data (solid points) from Ref.~\cite{alicespec}. The solid curve is Eq.~(\ref{alphas}) as defined in Ref.~\cite{alicetomspec}. The dashed curve is the solid curve reduced by factor 0.83. The open circles are derived in the present study from \pp\ $\bar p_t$ data in Ref.~\cite{alicempt}.
Right: Efficiency parameter $\xi$ vs cutoff parameter $p_{t,cut}$ defined as a running integral of the unit-normal TCM spectrum soft component $\hat S_0(p_t)$ with fixed slope parameter $T = 145$ MeV. Other curves are explained in the text.
 } 
\end{figure}

 Figure~\ref{params} (left) shows values for $\alpha(\sqrt{s})$ (solid points) obtained for 200 GeV and 13 TeV from Refs.~\cite{ppquad,alicespec} respectively. The solid curve is Eq.~(\ref{alphas}) above from Ref.~\cite{alicetomspec}  based on measured properties of isolated jets~\cite{jetspec2}. The open points correspond to $\bar p_t$ trends in the present study (described below). The dashed curve is the solid curve reduced by factor $\approx 0.83$ ($\approx 0.5/0.6$) corresponding to the reduced angular acceptance $\Delta \eta = 0.6$ in Ref.~\cite{alicempt} compared to previous results for $\Delta \eta = 2.0$~\cite{ppquad}.

  Figure~\ref{params} (right) shows the running integral $\int_{p_{t,cut}}^\infty dp_t p_t \hat S_0(p_t)$ (solid curve) representing the ratio $\xi = n'_s / n_s < 1$ of accepted to corrected soft-component multiplicities, the efficiency factor due to a \pt\ acceptance cutoff at some lower $p_{t,cut}$. In what follows primes denote uncorrected quantities responding to the \pt\ cutoff. That curve is independent of $\sqrt{s}$ or \nch\ given soft-component results in Ref.~\cite{alicetomspec}. The vertical hatched band marks the nominal acceptance cut for the analysis in Ref.~\cite{alicempt} corresponding to an expected acceptance fraction $\xi = 0.83$. 
Results from Ref.~\cite{alicetommpt} and the present study indicate that the effective acceptance cut for \pp\ data is actually closer to 0.17 GeV/c, and the measured acceptance fraction is in the range 0.76 - 0.80 indicated by the horizontal hatched band.
The dashed curve shows the strong increase of $\bar p_t'$ soft component $\bar p_{ts}'$ with increasing \pt\ cut, whereas the dash-dotted curve shows the corresponding decrease in the {\em total}-\pt\ soft component $\bar P'_{ts}$ (divided by the nominal NSD value of soft component $n_s$). The product $\bar P_{ts}' = n_s' \bar p_{ts}'$ is relatively insensitive to a \pt\ cut. In what follows it is assumed that $\bar P_{ts}'$ is a good approximation to the full-acceptance value $\bar P_{ts}$ (within 5\%). It is also assumed that hard components $n_h$ and $\bar P_{th}$ are unaffected by the typical $p_{t,cut} \approx 0.15$ GeV/c limit given Fig.~\ref{pp1} (right).


The 200 GeV result in Fig.~\ref{enrat3} (left) includes variation of TCM model $\hat H_0(y_t,n_{ch})$ for seven multiplicity classes (thin solid curves) derived in Ref.~\cite{alicetomspec} from high-statistics spectrum data~\cite{ppquad} and used below to predict  $\bar p_{th}(n_s,\sqrt{s})$ for two collision energies. \nch\ dependence of the TCM hard component is interpreted to reflect bias of the underlying jet energy spectrum by the imposed \nch\ condition.

\begin{figure}[h]
  \includegraphics[height=1.66in]{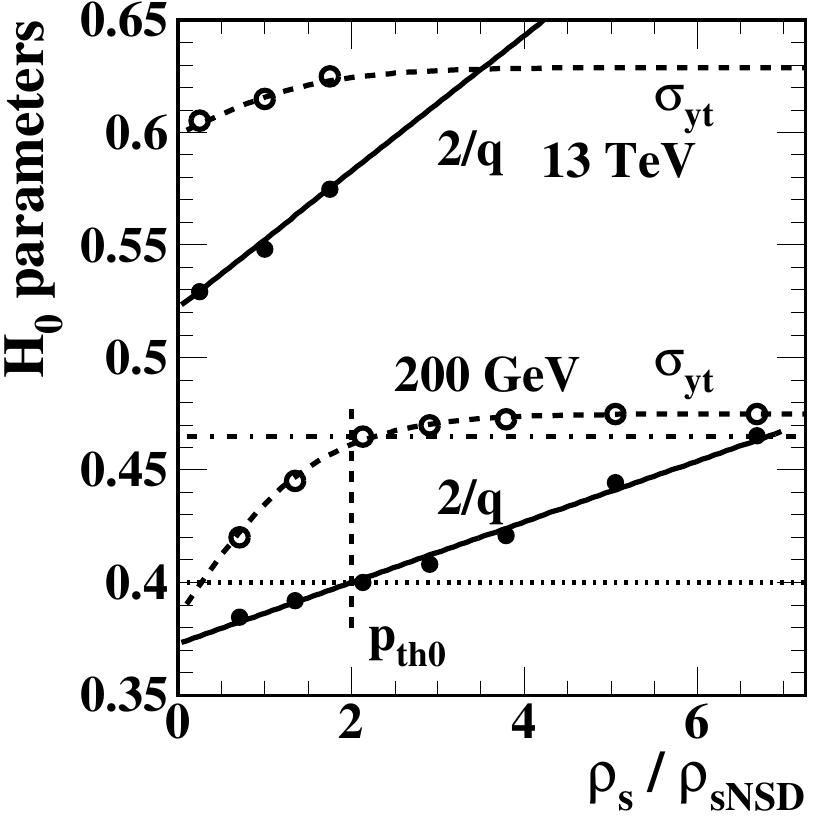}
   \includegraphics[height=1.63in]{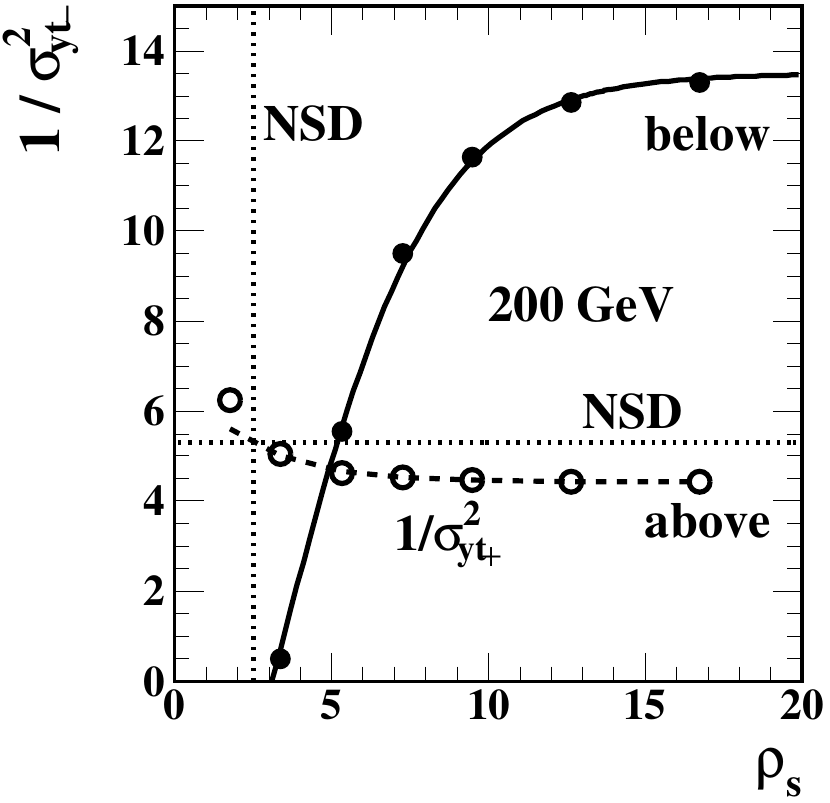}
\caption{\label{newparams1}
Left: Hard-component parameters varying with $n_{ch}'$ or $\bar \rho_s$. The solid and dashed curves through parameter data are as they appear in Ref.~\cite{alicetomspec}.  $\bar \rho_{s,ref} = 2.5$ for 200 GeV NSD \pp\ collisions and 6 for 13 TeV INEL $> 0$ (inelastic events with at least one charged particle accepted) collisions. The factor 2 in $2/q$ permits greater plot sensitivity.  
Right:  Variation of the Gaussian width below the hard-component mode $\sigma_{y_t-}$ (solid points) for multiplicity classes $n = 2$-7 of 200 GeV spectra that accommodates those data. The Gaussian width above the mode $\sigma_{y_t+}$ (open points) is included for comparison. Note that the Gaussian model is symmetric for $\bar \rho_s \approx 5 \approx 2 \bar \rho_{sNSD}$.
 }  
\end{figure}

Figure~\ref{newparams1} (left) shows TCM hard-component parameters $\sigma_{y_t+}$ (peak width above the mode) and $q$ (power-law exponent) as functions of soft-component charge-density ratio $\bar \rho_s / \bar \rho_{sNSD}$ for 200 GeV (lower) and 13 TeV (upper), where $\bar \rho_{sNSD}$ is the NSD soft density for each energy.

Figure~\ref{newparams1} (right) shows  TCM hard-component parameter $\sigma_{y_t-}$ (peak width below the mode) for 200 GeV. Note that for $\bar \rho_s   \approx 2 \bar \rho_{sNSD} \approx 5$, $\sigma_{y_t+}$ and $\sigma_{y_t-}$ are the same -- the hard-component peak (its Gaussian part) is approximately symmetric. The forms of the plotted variables facilitate modeling with simple functions.  These trends are extended to describe $\bar p_t(n_s,\sqrt{s})$ vs $\bar \rho_s$ trends at two energies in Sec.~\ref{nchvar} and App.~\ref{equations}.

\subsection{Updated $\bf \bar p_t$ TCM for p-p collisions} \label{ppmptupdate}

The quantities $\bar p_{th}(n_s,\sqrt{s})$, $\alpha(\sqrt{s})$ and the effective detector-acceptance ratio $\xi$ introduced above are used in this subsection to update the \mmpt\ TCM from Ref.~\cite{alicetommpt}. The TCM for charge yields integrated within some angular acceptance $2\pi$ and $\Delta \eta$ (e.g.\ 0.6 for Ref.~\cite{alicempt}) is 
\bea
n_{ch} &=& n_{s} + n_{h}
\\ \nonumber
&=& n_{s}[1+ x(n_s)],
\\ \nonumber
n_{ch}' / n_s&=& \xi+ x(n_s),
\eea
where $x(n_s) \equiv n_{h} / n_{s} \approx  \alpha \bar \rho_s$ is the ratio of hard-component to soft-component yields~\cite{ppprd} and $\alpha(\sqrt{s})$ is shown in Fig.~\ref{params} (left). The TCM for ensemble-mean integrated total $\bar P_t$ within acceptance $\Delta \eta$ from \pp\ collisions for given $(n_{ch},\sqrt{s})$ is simply expressed as
\bea \label{mptsimple}
\bar P_t &=& \bar P_{ts} + \bar P_{th}
\\ \nonumber
&=& n_s \bar p_{ts} + n_h \bar p_{th}.
\eea
The conventional intensive ratio of extensive quantities
\bea \label{ppmpttcm}
\bar p_t' \equiv \frac{\bar P_t'} {n_{ch}'} &\approx & \frac{\bar p_{ts} + x(n_s) \bar p_{th}(n_s)}{\xi + x(n_s)}
\eea
(assuming $\bar P_t' \approx \bar P_t$) conflates two simple TCM trends and in effect partially cancels MB dijet manifestations apparent in the form of $x(n_s)$.  Alternatively, the ratio
\bea \label{niceeq}
\frac{n_{ch}'}{n_s} \bar p_t'   \approx \frac{ \bar P_t}{n_s} &= & \bar p_{ts} + x(n_s) \bar p_{th}(n_s)
\\ \nonumber
&\approx& \bar p_{ts} + \alpha(\sqrt{s})\, \bar \rho_s \, \bar p_{th}(n_s,\sqrt{s})
\eea
preserves the simplicity of Eq.~(\ref{mptsimple}) and provides a convenient basis for testing the TCM hypothesis precisely.

\begin{figure}[h]
   \includegraphics[width=1.65in,height=1.6in]{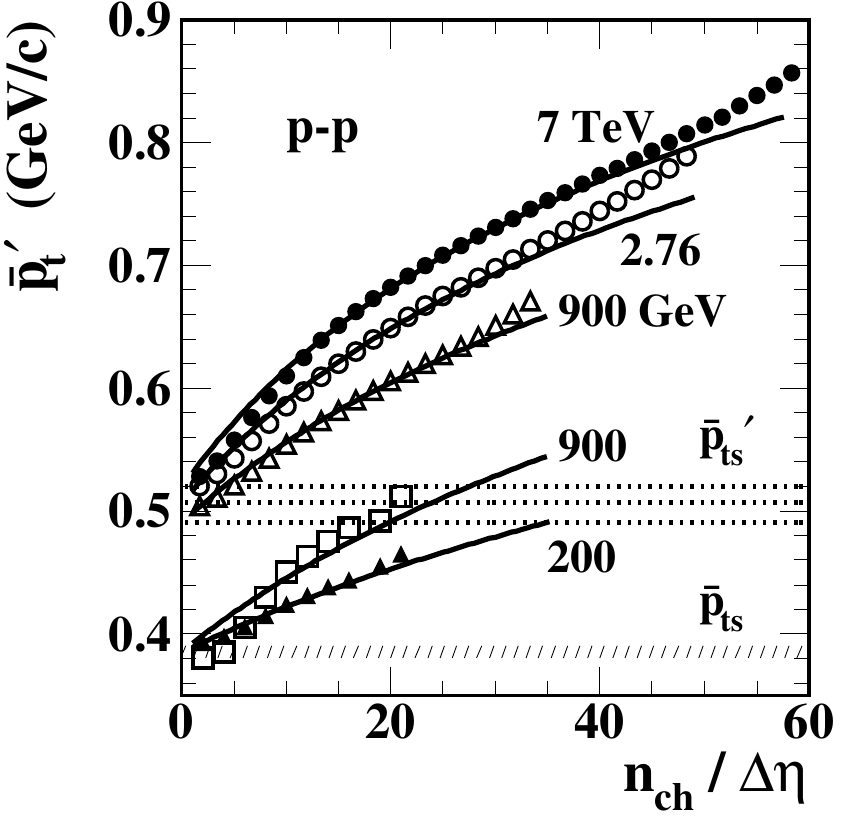}
   \includegraphics[width=1.65in,height=1.6in]{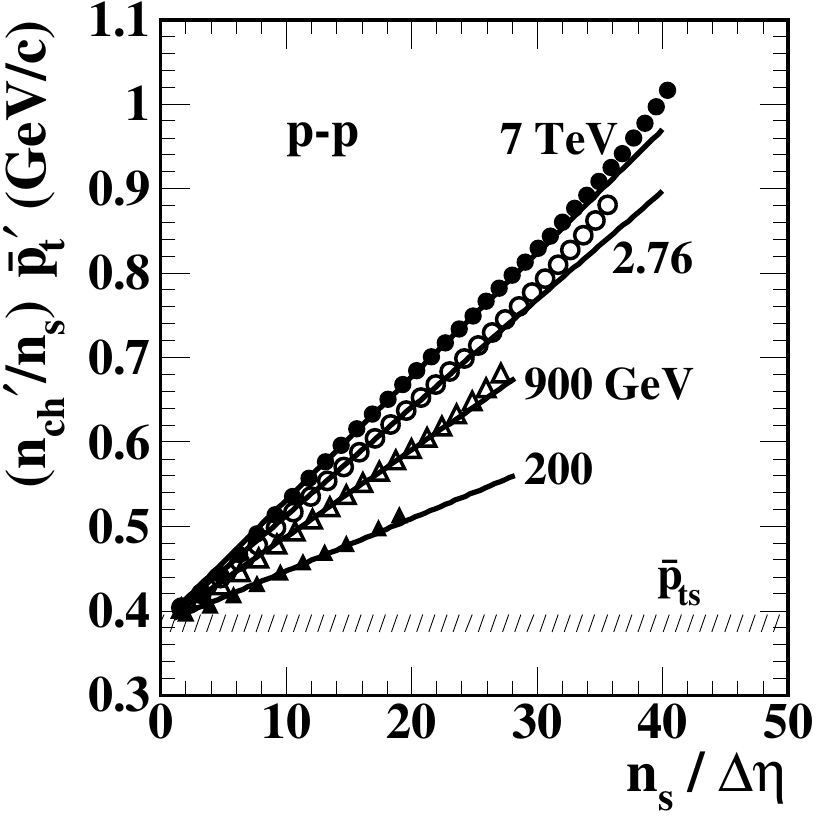}
 \caption{\label{alice5a}
 Left: \mmpt\ vs \nch\ for several collision energies. The upper group of points is from Ref.~\cite{alicempt}. The lower 900 GeV data from UA1 derived from a ``power-law'' spectrum model~\cite{ua1mpt} fall significantly above the TCM for that energy (solid curve) but are consistent with the TCM form with amplitude adjusted. The 200 GeV STAR data are spectrum integrals from Ref.~\cite{ppprd}.
Right: Data from the left panel multiplied by factor ${n'_{ch}} / n_s$ that removes a jet contribution to denominator ${n'_{ch}}$ of $\bar p_t'$ and the effect of a \pt\ cut on its soft component. 
} 
  \end{figure}

Figure~\ref{alice5a} (left) shows $\bar p_t$ data for four \pp\ collision energies from 
the RHIC  (solid triangles~\cite{ppprd}),
the Sp\=pS  (open boxes~\cite{ua1mpt})%
\footnote{The analysis of Ref.~\cite{ua1mpt} inferred \mmpt\ values by fitting a ``power-law'' model function to \pt\ spectra. Possible biases arising from that method, especially at lower \nch, are discussed in Ref.~\cite{ppprd}.}%
and the LHC (upper points~\cite{alicempt}) increasing monotonically with charge density $\bar \rho_0 = n_{ch} / \Delta \eta$. The lower points and curves correspond to full \pt\ acceptance.  For acceptance extending down to zero ($\xi = 1$), $\bar p_t' \rightarrow \bar p_t$ in Eq.~(\ref{ppmpttcm}) should vary between the universal lower limit $\bar p_{ts} \approx 0.40$ GeV/c ($n_{ch} \rightarrow 0$) and  $\approx \bar p_{th0}$ ($n_{ch} \rightarrow \infty$) as limiting cases. For a lower-\pt\ cut $p_{t,cut} > 0$ the lower limit is $\bar p_{ts}' = \bar p_{ts} / \xi$ (upper dotted lines) and the data are systematically shifted upward (upper points and curves).  Solid curves represent the \pp\ $\bar p_t$ TCM from this study.

\begin{figure}[h]
   \includegraphics[width=1.65in,height=1.6in]{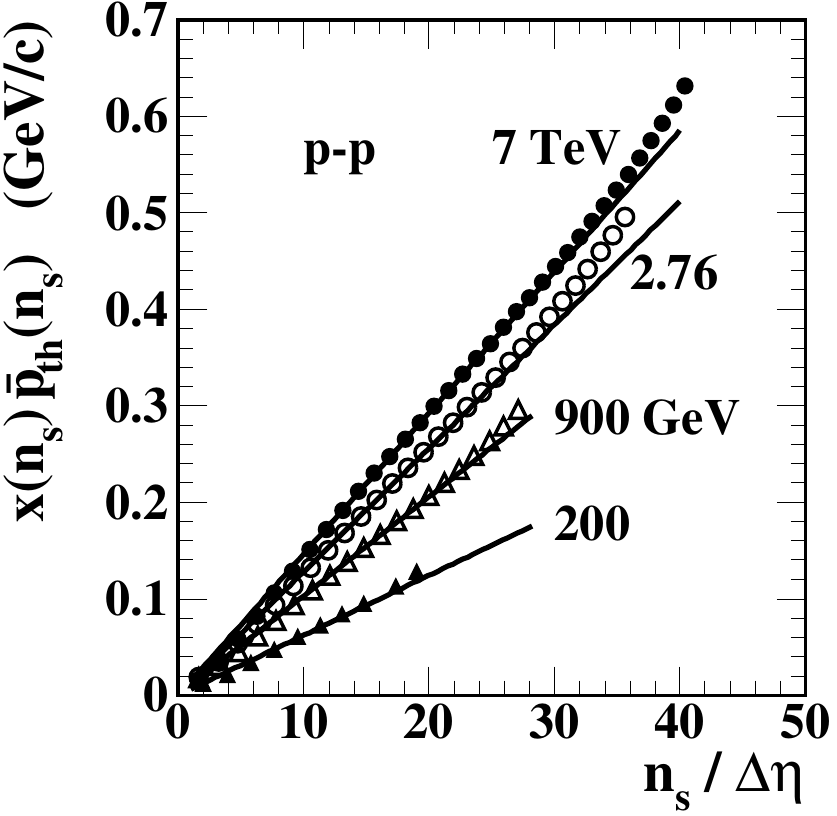}
   \includegraphics[width=1.65in,height=1.6in]{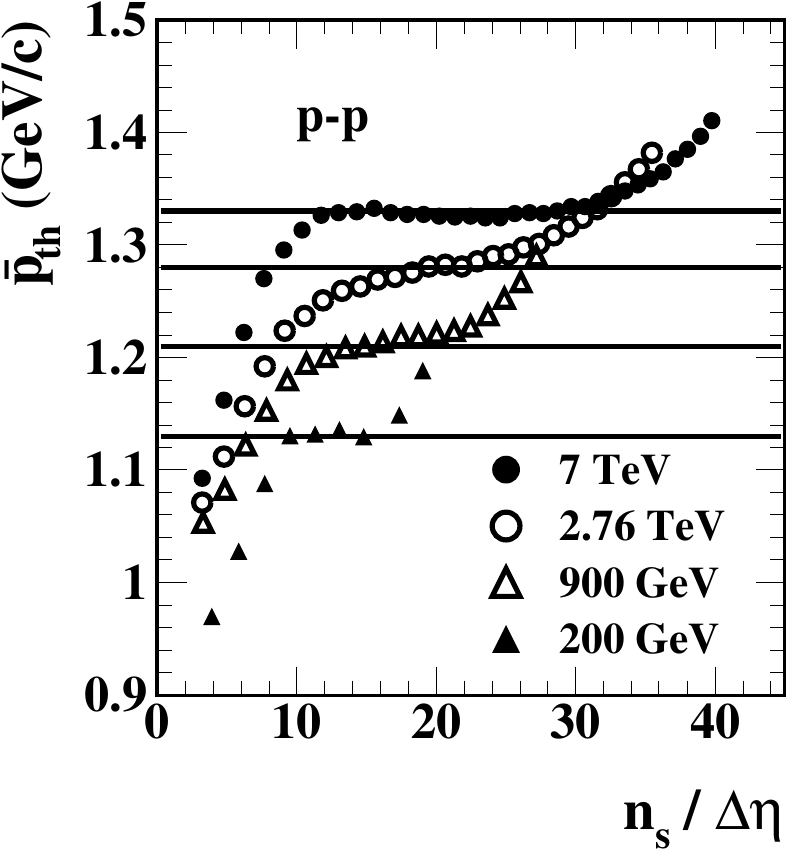}
 \caption{\label{alice5}
Left: 
Subtracting a universal soft component $\bar p_{ts}$ from data in Fig.~\ref{alice5a} (right) isolates product $x(n_s,\sqrt{s}) \bar p_{th}(n_s,\sqrt{s})$ according to Eq.~(\ref{niceeq}).
 Right: Hard components  $\bar p_{th}(n_s)$ (points) isolated from data at left per Eq.~(\ref{niceeq}). The lines represent mean values $\bar p_{th0}$ extracted from \pp\ spectra (Fig.~\ref{enrat3}, right). The  $\bar p_{th}(n_s)$ data vary significantly about the mean values as expected from results in Ref.~\cite{alicetomspec} in the next subsection.
 } 
  \end{figure}

Figure~\ref{alice5a} (right) shows data on the left transformed via Eq.~(\ref{niceeq}) to $(n_{ch}' / n_s) \bar p_t' \approx  \bar P_t / n_s$ (points). The TCM curves undergo the same transformation and the slopes of the resulting lines are $\alpha(\sqrt{s}) \bar p_{th0}(\sqrt{s})$.
Figure~\ref{alice5} (left) shows quantity $(n_{ch}' / n_s) \bar p_t' - \bar p_{ts} \approx x(n_s,\sqrt{s}) \bar p_{th}(n_s,\sqrt{s})$ (points). The TCM lines are $ \alpha(\sqrt{s}) \, \bar \rho_s\, \bar p_{th0}$; the two factors multiplying $\bar \rho_s$ are derived from \pt\ spectrum data. 

Figure~\ref{alice5} (right) shows data in the left panel divided by $\alpha(\sqrt{s}) \, \bar \rho_s$ to obtain $\bar p_{th}(n_s,\sqrt{s})$ vs $\bar \rho_s$.  The solid lines represent the $\bar p_{th0}$ values (open circles) in Fig.~\ref{enrat3} (right) derived from TCM spectrum hard components. Values for $\alpha(\sqrt{s})$ plotted as open circles in Fig.~\ref{params} (left) are consistent with correspondence between $\bar p_{th}(n_s,\sqrt{s})$ data (points) and $\bar p_{th0}$ values (lines) as shown.  The $\alpha(\sqrt{s})$ values for various energies and detector systems are quantitatively consistent to a few percent given the differences in acceptance $\Delta \eta$ between experiments. 
Some \nch\ dependence of $\bar p_{th}$ mean values was already expected based on results from Refs.~\cite{ppprd,alicetomspec} [see the corresponding 200 GeV hard-component \nch\ dependence (thin solid curves) in Fig.~\ref{enrat3} (left)]. Note that there is substantial systematic disagreement between Figure~\ref{alice5} (right) and the comparable figure in Ref.~\cite{alicetommpt} due to the incorrect assumption in the earlier study that $\alpha$ is not energy dependent (the 200 GeV $\alpha$ value 0.006 was retained for all collision energies).

\subsection{$\bf n_{ch}$-dependent TCM hard component}  \label{nchvar}

New information about the \nch\ dependence of the TCM spectrum hard component reported in Ref.~\cite{alicetomspec} can be applied to $\bar p_{th}(n_s)$ results in Fig.~\ref{alice5} (right). The $\bar p_{th}(n_s)$ data for limiting cases 200 GeV and 7 TeV \pp\ collisions are considered here. Details of the corresponding hard-component model parameters are provided in App.~\ref{equations}.

The TCM spectrum hard component on transverse rapidity \yt\ is modeled by a Gaussian with exponential tail (equivalent to a power-law trend on \pt). The transition point from Gaussian to exponential is determined by slope matching and depends therefore on the model parameters. The hard-component model shape is determined by separate widths $\sigma_{y_t+}$ and $\sigma_{y_t-}$ above and below the peak mode and exponential parameter $q$, all of which vary with control parameter $\bar \rho_s = n_s / \Delta \eta$ to accommodate spectrum data as established in Ref.~\cite{alicetomspec}. For $\bar \rho_s < 15$ the $\bar p_{th}(n_s)$ trend is dominated by $\sigma_{y_t-}(n_s)$. For $\bar \rho_s > 15$ the trend is dominated by $\sigma_{y_t+}(n_s)$ and $q(n_s)$. For higher energies only $q(n_s)$ matters because the Gaussian-exponential transition is close to the mode.

\begin{figure}[h]
    \includegraphics[width=3.3in,height=1.6in]{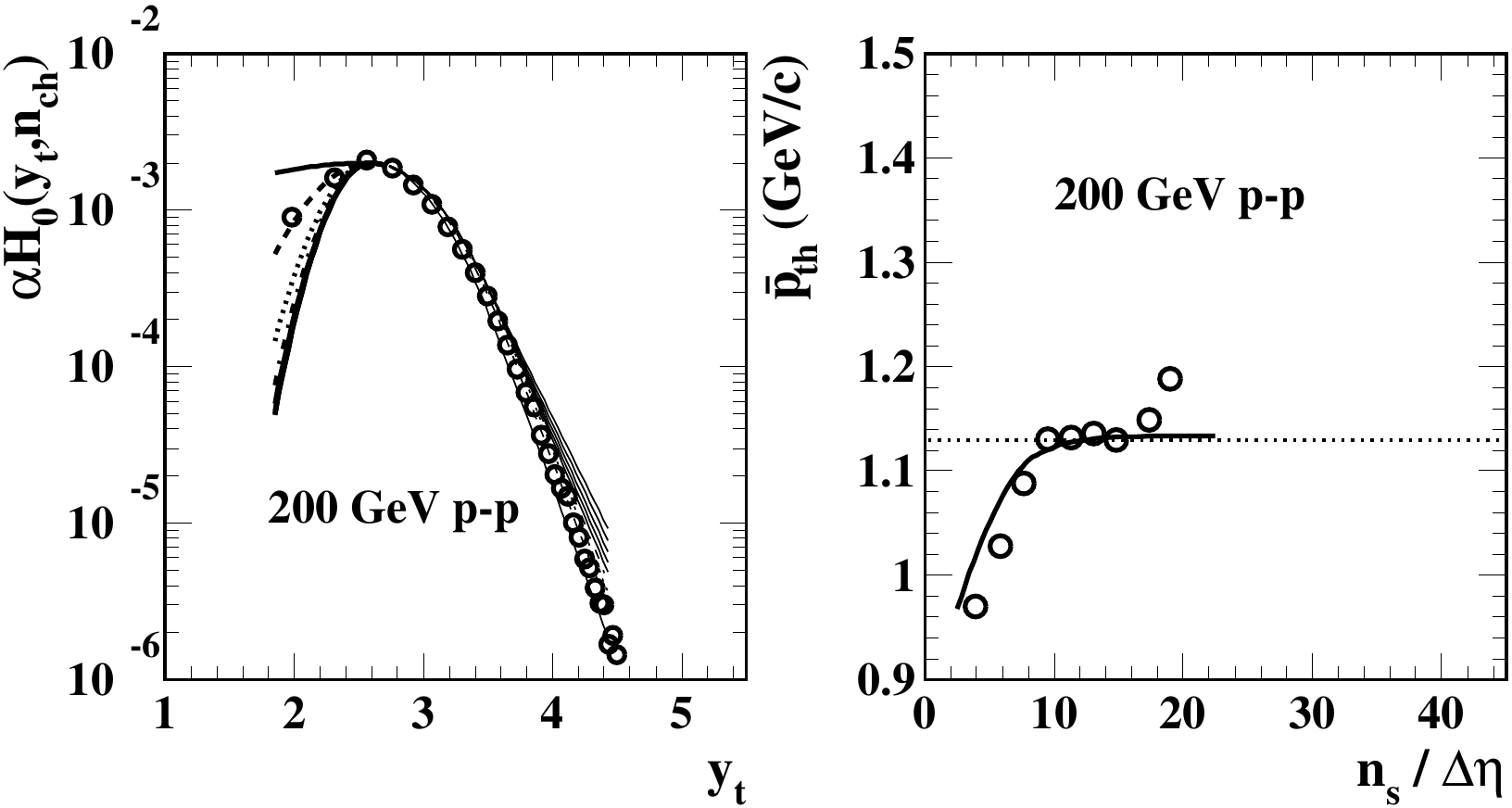}
 \caption{\label{300b}
Left: \pt\ spectrum hard-component models for 200 GeV \pp\ collisions and a range of \nch\ based on a TCM parametrization from Ref.~\cite{alicetomspec} reviewed in App.~\ref{equations}.
 Right: $\bar p_{th}(n_s)$ values inferred from 200 GeV  \pp\ \mmpt\ data appearing in Fig.~\ref{alice5} (right) (points) compared to a $\bar p_{th}(n_s)$ trend (curve) inferred from the model functions in the left panel.
 } 
  \end{figure}

Figure~\ref{300b} (left) shows 200 GeV TCM hard-component models for nine \pp\ multiplicity classes based on corresponding parameter variations in App.~\ref{equations}. The actual parameter values are determined by expressions provided in App.~\ref{200gevparams}. The model functions accurately represent the spectrum hard components in Fig.~\ref{pp1} (right) of this study repeated from Ref.~\cite{ppquad}. Estimates for $\bar p_{th}(n_s)$ can then be obtained by integrating each model function.

Figure~\ref{300b} (right) shows the 200 GeV $\bar p_{th}(n_s)$ data from Fig.~\ref{alice5} (right) (points). The curve is determined by the  $\bar p_{th}(n_s)$ values obtained from the model functions in the left panel. The correspondence between data values and TCM curve is good and illustrates the accuracy of the TCM description of 200 GeV spectrum data in Ref.~\cite{alicetomspec}.

\begin{figure}[h]
   \includegraphics[width=3.3in,height=1.6in]{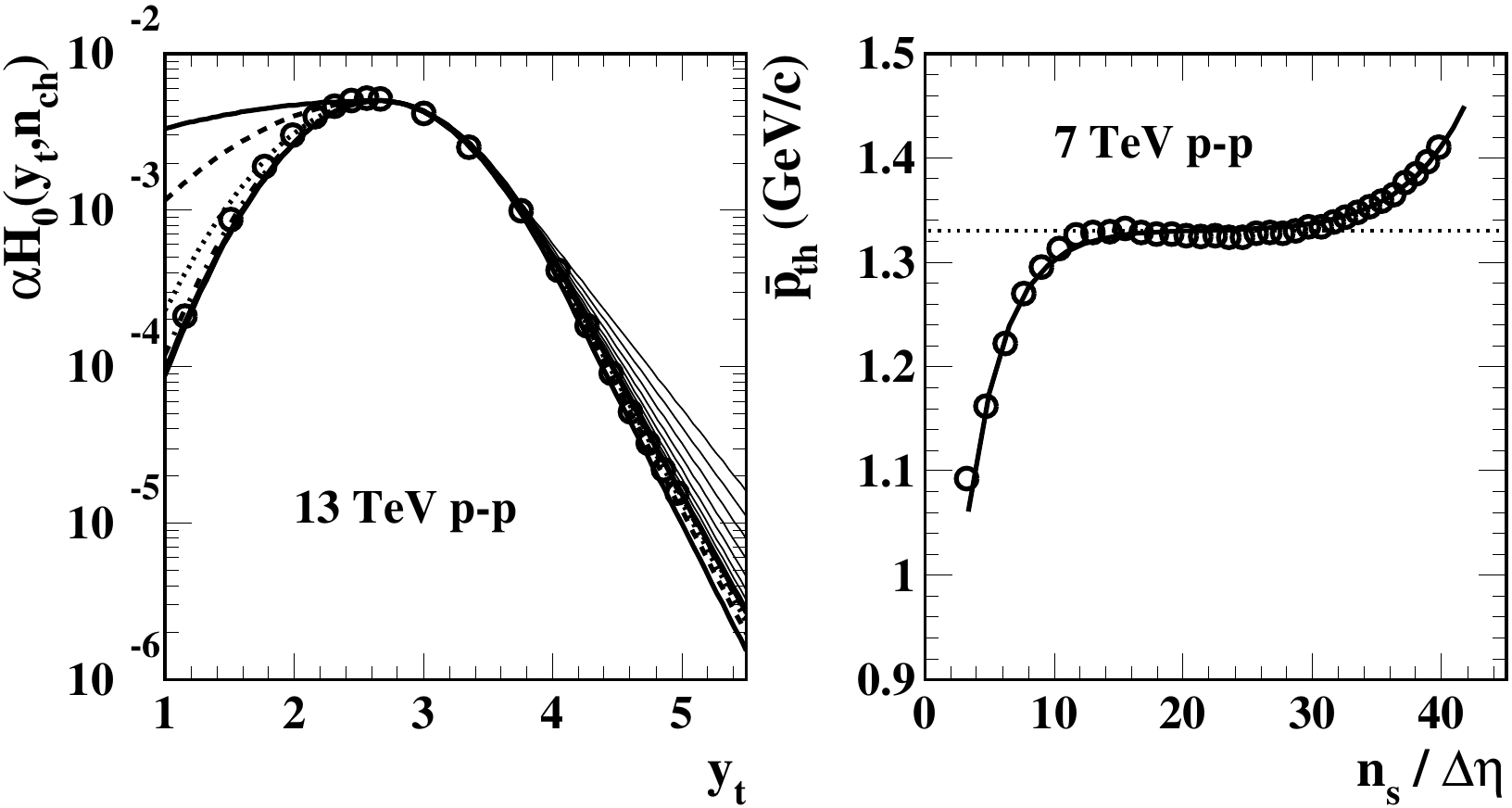}
   \caption{\label{300a}
Left: \pt\ spectrum hard-component models for 13 TeV \pp\ collisions and a range of \nch\ based on a TCM parametrization from Ref.~\cite{alicetomspec} reviewed in App.~\ref{equations} compared to spectrum data from Ref.~\cite{alicespec} for inelastic \pp\ collisions (points).
 Right: $\bar p_{th}(n_s)$ values inferred from 7 TeV  \pp\ \mmpt\ data appearing in Fig.~\ref{alice5} (right) (points) compared to a $\bar p_{th}(n_s)$ trend (curve) inferred from the model functions in the left panel.
 } 
  \end{figure}

Figure~\ref{300a} (left) shows  13 TeV TCM hard-component models for several \pp\ multiplicity classes.  The parameter variation is determined by expressions provided in App.~\ref{7tevparams}. Those expressions interpolate from 13 TeV as in Fig.~\ref{newparams2} (left) to 7 TeV and extrapolate on $\bar \rho_s$ from the rather limited 13 TeV multiplicity range in Ref.~\cite{alicespec}.

Figure~\ref{300a} (right) shows 7 TeV $\bar p_{th}(n_s)$ data from Fig.~\ref{alice5} (right) (points). The curve is determined by the  $\bar p_{th}(n_s)$ values obtained from spectrum model functions at left. Whereas the width above the mode $\sigma_{y_t+}$ is just as determined in Ref.~\cite{alicetomspec} power-law exponent $q$ is freely fitted to accommodate $\bar p_{th}$ data for $\bar \rho_s > 15$. The form of width below the mode $\sigma_{y_t-}(n_s)$ (for which there is no information from spectrum data) is maintained the same as that for 200 GeV, but the amplitude is increased by factor 2 to accommodate the 7 TeV $\bar p_{th}$ data for $\bar \rho_s < 15$. The correspondence between \mmpt\ data and TCM is again good. In turn there is a quantitative correspondence between \pp\ \pt\  spectrum hard components and jet properties accurate at the percent level. These detailed TCM results further buttress the conclusion that ensemble-mean $\bar p_t$ variation is essentially completely determined by MB dijets.

\subsection{p-p TCM conclusions}

\pp\ collision data reveal that there are two principal sources of hadrons: (a) fragmentation of low-$x$ gluons from dissociated projectile nucleons (soft component $\bar \rho_s$) that emerge outside the collision space-time volume and thus do not rescatter ~\cite{bialas,witpa,witpa2,witpa3} and (b) jet fragments from large-angle-scattered low-$x$ gluons (hard component $\bar \rho_h$). The two are simply related to results from late-seventies fixed-target $h$-A experiments (soft)~\cite{witpa} and extensive jet measurements over the past thirty years (hard)~\cite{jetspec2}. The resulting TCM quantitatively describes all aspects of \pp\ collisions including yields~\cite{alicetomspec}, spectra~\cite{ppprd,ppquad} and correlations~\cite{porter2,porter3,ppquad} (except a nonjet quadrupole~\cite{ppquad}). The present study is fully consistent with that statement.

The hard-component angular density $\bar \rho_{h} \approx \alpha(\sqrt{s})\, \bar \rho_s^2$ represents the dijet fragment density determined precisely by soft component $\bar \rho_s$. The \pt\ spectrum TCM hard component and underlying jet energy spectrum evolve according to the same rules~\cite{jetspec2}. Noneikonal: each participant gluon in one proton can interact with {\em any} participant gluon in the partner proton. The noneikonal $\bar \rho_h \propto \bar \rho_s^2$ trend for \pp\ collisions is the same for jet-related angular correlations~\cite{ppquad}, inferred spectrum hard component~\cite{ppprd,ppquad} and $\bar p_t$ hard component~\cite{alicetommpt}. The noneikonal trend implies that for large \nch\ \pp\ collisions are very jetty: for $\bar \rho_s\approx 10\, \bar \rho_{sNSD}$ -- implying a 100-fold increase in dijet production -- hard/soft ratio $x$ is 0.15 at 200 GeV and 0.72 at 7 TeV. It also implies that the concepts of centrality and impact parameter do not apply to \pp\ collisions, as confirmed by measurements of a NJ quadrupole component of 2D angular correlations (no eccentricity dependence is evident)~\cite{ppquad}. Instead, \nch\ is determined eventwise mainly by the  fluctuating depth of the splitting cascade on momentum fraction $x$ and consequent production of low-$x$ gluons within a projectile proton.

The \pt-spectrum hard component is quantitatively related to measured properties of isolated jets~\cite{fragevo,ppquad}. An \nch\ condition is observed to bias the underlying jet spectrum and thus the spectrum hard component~\cite{alicetomspec} as in Fig.~\ref{pp1} (right). Reference~\cite{alicetommpt} and the present study establish a direct connection between \mmpt\ hard component $\bar p_{th}(n_s)$ and spectrum hard component $H(p_t,n_s)$. That connection is buttressed by observation of similar variation of $\bar p_{th}(n_s)$ with \nch\ as in Figs.~\ref{300b} and \ref{300a}. The jet-related $n_s$ dependence may correspond to the ``change in slope'' referred to in Ref.~\cite{alicempt} (in re its Fig.~1). The energy dependence of $\bar p_{th0}(\sqrt{s})$ as in Fig.~\ref{enrat3} (right) also corresponds to isolated-jet properties as in Ref.~\cite{jetspec2}. Thus, a variety of \pp\ data provide overwhelming evidence that MB dijets dominate \pp\ collisions and $\bar p_t(n_{ch},\sqrt{s})$ trends.

\section{$\bf \bar p_t$ TCM for $\bf p$-$\bf Pb$  collisions} \label{ppb}

The TCM for \aa\ collisions relies on participant-pair number $N_{part}/2$, number of \nn\ binary collisions $N_{bin}$ and mean number of binary collisions per participant pair $\nu \equiv 2N_{bin} / N_{part}$. As noted in Ref.~\cite{alicetommpt} the $\bar p_t$ trend for \ppb\ collisions is intermediate between \pp\ (at lower \nch) and \pbpb\ (at higher \nch) suggesting a formulation of the TCM for the \ppb\ collision system based on the product $x(n_s) \nu(n_s)$. For \pp\ collisions $\nu \equiv 1$ and $x(n_s) \approx \alpha \bar \rho_s$ based on spectrum studies~\cite{ppprd,ppquad}. For \aa\ collisions $\nu$ has been defined by a Glauber Monte Carlo~\cite{powerlaw} and $x(\nu)$ has been inferred from the trend of per-participant hadron yields as in Fig.~\ref{nchaa} below. The basis for a \pa\ TCM is an intermediate generalization of $x(n_s) \nu(n_s)$.

\subsection{Formulating a $\bf p$-Pb TCM} \label{patcm}

A \ppb\ TCM can be established  based on the product $x(n_s)\nu(n_s)$ with soft multiplicity $n_s$ ($\propto$ total number of {\em participant} low-$x$ gluons) as the independent variable for all collision systems. A general TCM for  {\em extensive} variable $Z$ (e.g.\ total $P_t$ integrated within some angular acceptance) applicable to any A-B collision system can be expressed as
\bea \label{xeqn}
Z &=& Z_s + Z_h
\\ \nonumber
&=& \frac{N_{part}}{2} n_{sNN}\, z_{sNN} + N_{bin} n_{hNN}\, z_{hNN}
\eea
representing factorization between A-B geometry parameters and mean \nn\ hadron production within the A-B system. For charge multiplicity the TCM relation is
\bea \label{nchppb}
n_{ch} &=& n_s + n_h
\\ \nonumber
&=& \frac{N_{part}}{2} n_{sNN}(n_s) + N_{bin} n_{hNN}(n_s)
\\ \nonumber
\frac{2}{N_{part}}  n_{ch} &=& n_{sNN}(n_s)  \left[1 + x(n_s)\nu(n_s) \right]
\\ \nonumber
\frac{n_{ch}}{n_{s}} &=& 1 + x(n_s)\nu(n_s) 
\\ \nonumber
\frac{n_{ch}'}{n_{s}} &=& \xi + x(n_s)\nu(n_s),
\eea
where  \pp, \pa\ and \aa\ are possible collision systems, $x(n_s) \equiv n_{hNN} / n_{sNN} \approx \alpha \bar \rho_{sNN}$ (for \pp\ and \pa), $n_{s} = [N_{part}(n_s)/2]\, n_{sNN}(n_s)$ is a factorized soft-component yield for any system ($N_{part} / 2 \equiv 1$ for \pp) and $n_h(n_s) = N_{bin}(n_s) n_{hNN}(n_s)$ is a factorized hard component.   

\
The corresponding TCM for $\bar p_t'$ with nonzero $p_{t,cut}$ is
\bea \label{pampttcm}
\bar p_t' \equiv \frac{\bar P_t'} {n_{ch}'} &\approx& \frac{\bar p_{ts} + x(n_s) \nu(n_s) \, \bar p_{thNN}(n_s)}{\xi + x(n_s)\, \nu(n_s)}
\\ \nonumber
\frac{n_{ch}'}{n_{s}} \bar p_t' &\approx& \frac{\bar P_t}{n_{s}}
= \bar p_{ts} + x(n_s)\nu(n_s) \, \bar p_{thNN}(n_s),
\eea
where $\bar p_{tsNN} \rightarrow \bar p_{ts} \approx 0.4$ GeV/c for all systems.
Those expressions go to the \pp\ equivalent as described in Sec.~\ref{ppmptupdate} when $\nu(n_s) \rightarrow 1$. \pa\ data require evolution of factors $x(n_s)\, \nu(n_s)$ from \pp-like to \aa-like behavior near a transition point  $\bar \rho_{s0}$, and it is assumed that  $\bar p_{thNN}(n_s) \rightarrow \bar p_{th0}$ retains a fixed \pp\ (\nn) value in the \pa\ system (no jet modification in small systems).


Soft density $\bar \rho_s$, representing (by hypothesis) participant low-$x$ gluons, is observed to be the common TCM parameter for all collision systems. The factorization
\bea \label{pafactor}
\bar \rho_s &=&\bar \rho_{sNN}(n_s) \,  N_{part}(n_s)/2
\eea
then defines $\bar \rho_{sNN}(n_s)$ for any collision system wherein $N_{part}(n_s)/2$ is defined. Hard/soft yields are related by parameter $x(n_s) \equiv n_{hNN}(n_s) / n_{sNN}(n_s)$. Collision ``geometry'' is represented by $\nu(n_s) \equiv 2 N_{bin}(n_s)/N_{part}(n_s)$. (See Eq.~(\ref{xnuu}) and surrounding text for further details.) For \pp\ collisions $x(n_s) \approx \alpha \bar \rho_{s}$ represents a noneikonal collision system~\cite{ppprd,ppquad}. More generally $x(n_s) \approx \alpha \bar \rho_{sNN}(n_s)$ for \pn\ collisions within \pa\ collisions.  It follows that $N_{part}(n_s)/2 = \alpha \bar \rho_s / x(n_s)$ assuming some model for $x(n_s)$. For \pa\ collisions $N_{part}$, $N_{bin} = N_{part} - 1$, and $\nu(n_s) \equiv 2N_{bin} / N_{part}$ are then all determined by $x(n_s)$ within a self-consistent \pa\ TCM. 

Based on previous analysis in Ref.~\cite{alicetommpt} \ppb\ data indicate that \mmpt\ increases with \nch\ according to a \pp\ trend for lower \nch\ but less rapidly above a transition point. That behavior suggests a similar structure for $x(n_s)$. The simplest case would be linear increase of $x(n_s)$ with $\bar \rho_s$ also above the  transition point but with a reduced slope. 

\begin{figure}[h]
  \includegraphics[width=1.65in]{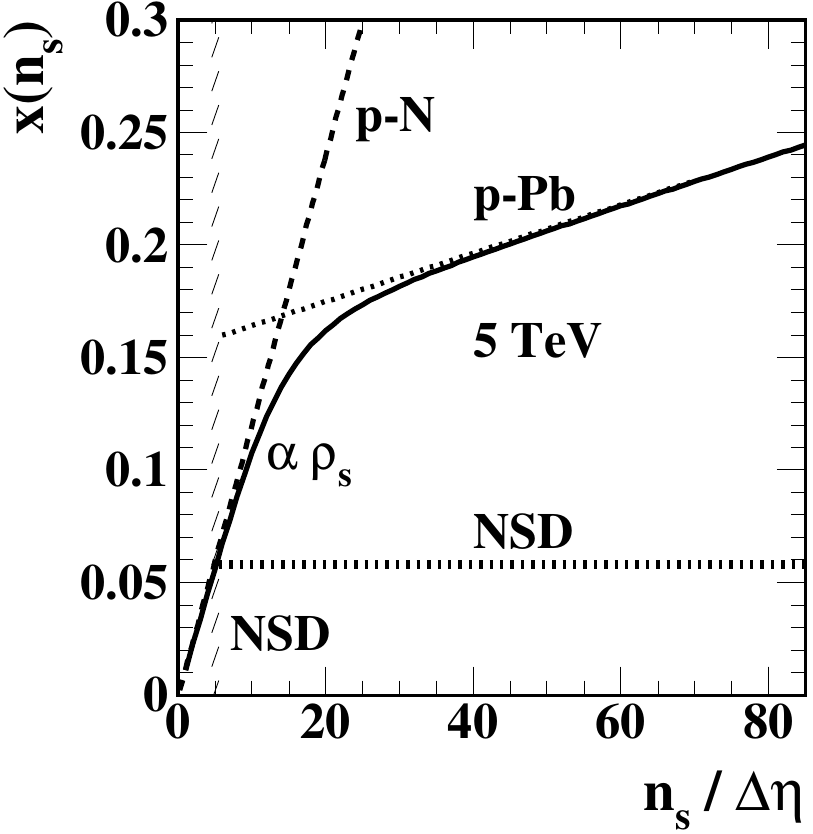}
  \includegraphics[width=1.65in]{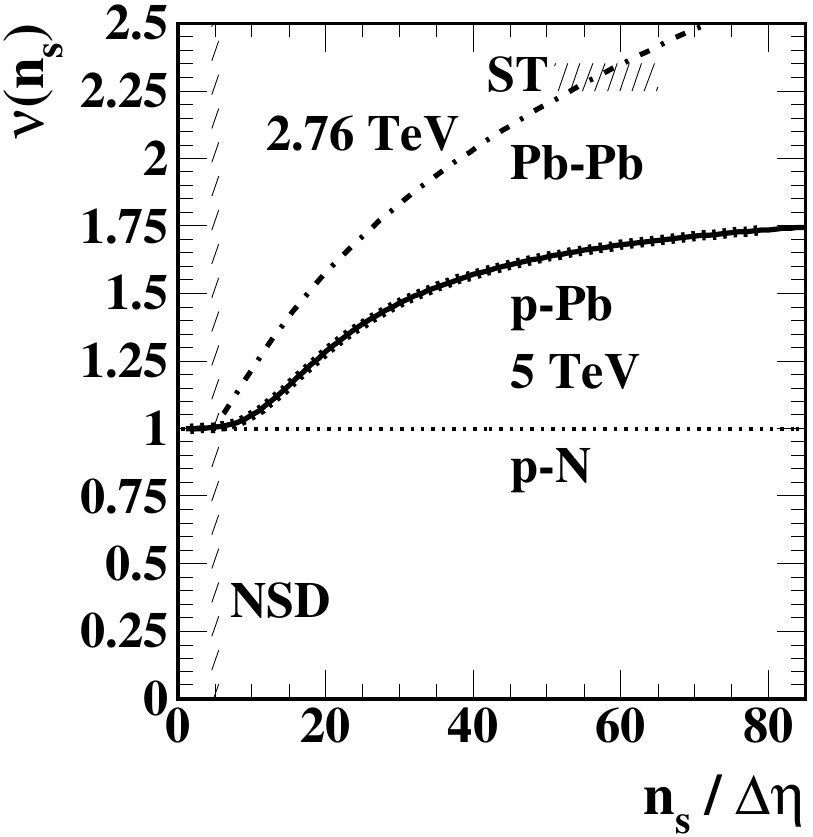}
  \caption{\label{paforms}
  Left: TCM parameter $x(n_s)$ vs $\bar \rho_s = n_s / \Delta \eta$ for 5 TeV \ppb\ collisions (solid curve). The trend varies from noneikonal $\alpha \bar \rho_s$ (dashed line)  to a continuing linear trend with 10-fold reduced slope (dotted line).
  Right: TCM mean-pathlength parameter $\nu(n_s)$ (solid curve) derived from the $x(n_s)$ trend in the left panel (see text). The dash-dotted curve is $\nu(n_s)$ for 2.76 TeV \pbpb\ collisions for comparison.
   } 
 \end{figure}

Figure~\ref{paforms} (left) shows a model for $x(n_s)$ in the form
\bea \label{xmodel}
x(n_s) &=& \frac{1}{\left\{[1/\alpha \bar \rho_s]^{n_1} + [1/f(n_s)]^{n_1}\right\}^{1/n_1}},
\eea
where $f(n_s) = \alpha [\bar \rho_{s0} + m_0(\bar \rho_s - \bar \rho_{s0})]$.
Below the transition at $\bar \rho_{s0}$, $x(n_s) = \alpha \bar \rho_s$ as for \pp\ collisions (dashed line). Above the transition $x(n_s)$ varies with slope $m_0 < 1$ (dotted line). Exponent $n_1$ controls the transition width. Specific parameter values for $x(n_s)$ are inferred below.

Figure~\ref{paforms} (right) shows $\nu \equiv 2N_{bin} / N_{part}$ for 5 TeV \ppb\ data (solid curve) based on $N_{part}(n_s)/2 = \alpha \bar \rho_s / x(n_s)$ and $N_{bin} = N_{part} - 1$ as noted above, with $x(n_s)$ as described in the left panel (solid curve). Within the TCM $n_s$ or $\bar \rho_s$ is the independent variable and other parameters are defined relative to it. However, measured data are defined relative to observed $n_{ch}'$ or corrected $n_{ch}$. The corresponding data $n_s$ must then be inferred by inverting Eqs.~(\ref{nchppb}) via linear interpolation. The bold dotted curve in the right panel is $\nu(n_s)$ derived for data \nch\ values. Congruence with the solid curve indicates exact agreement. The dash-dotted curve presents a $\nu  \approx (N_{part}/2)^{1/3} \approx (\bar \rho_s / \bar \rho_{sNSD})^{1/3}$ trend for 2.76 TeV \pbpb\ collisions for comparison, consistent with the eikonal approximation assumed for the Glauber model. Note that $\nu \in [1,6]$ for \pbpb\ but $\nu \in [1,2]$ for \ppb.


The trends in Fig.~\ref{paforms} are consistent with the following scenario: Increase of jet-related hadron production in \pa\ collisions may proceed via two mechanisms depending on control parameter $\bar \rho_s$: (a) increasing depth of splitting cascades on momentum fraction $x$ within single \pn\ collision partners that increases $\bar \rho_{sNN}(n_s) \approx \bar \rho_s$ with peripheral \pa\ geometry or (b) increasing participant-nucleon number $N_{part}(n_s)$ with increasing \pa\ centrality and $\bar \rho_{sNN}(n_s) < \bar \rho_s$. The relative contributions depend on probabilities. 
Below transition point $\bar \rho_{s0}$ single \pn\ collisions dominate and the noneikonal $\bar \rho_{hNN} \propto \bar \rho_{sNN}^2$  trend for dijet production observed in \pp\ collisions~\cite{ppquad} is determining.   Above $\bar \rho_{s0}$ \pa\ centrality dominates and increasing \pn\ binary-collision measure $\nu$ plays the determining role in measuring dijet production.

\subsection{TCM description of p-A $\bf \bar p_t'$ data}

\begin{figure}[h]
  \includegraphics[width=1.66in]{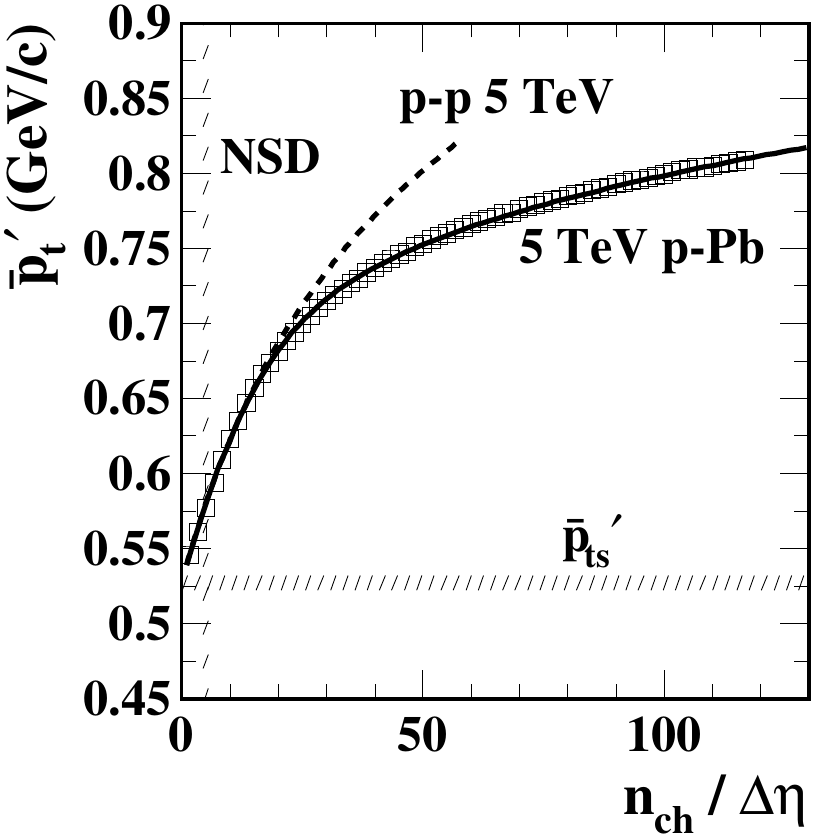}
  \includegraphics[width=1.64in]{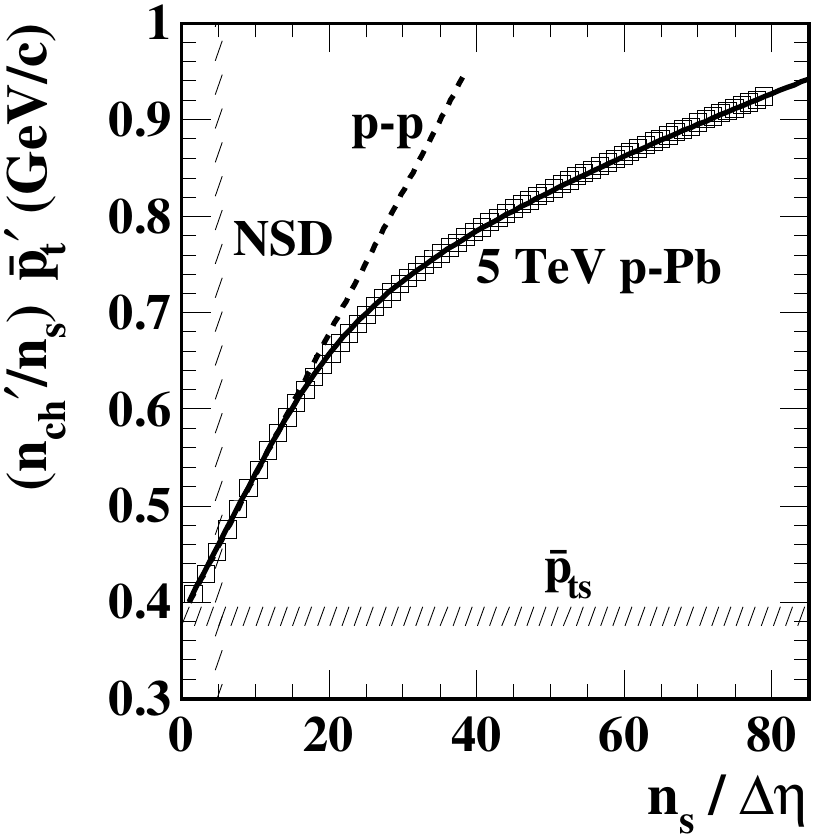}
  \caption{\label{padata}
  Left: Uncorrected $\bar p_{t}'$ data (points) vs corrected $\bar \rho_0 = n_{ch} / \Delta \eta$ for 5 TeV \ppb\ collisions from Ref.~\cite{alicempt}. The dashed curve is the \pp\ \mmpt\ TCM from Sec.~\ref{pptcmm}. The solid curve is the TCM for 5 TeV \pbpb\ collisions derived in this section.
  Right: Data from the left panel transformed to $(n_{ch}' / n_s)\, \bar p_t' \approx \bar P_t / n_{s}$ (points). The \pp\ trend is a straight line in this format.
   }  
\end{figure}

Figure~\ref{padata} (left) shows $\bar p_t'$ data for 5 TeV \ppb\ collisions vs corrected \nch\ (points) from Ref.~\cite{alicempt}. The dashed curve is the TCM for 5 TeV \pp\ collisions defined by Eq.~(\ref{ppmpttcm}) with $\alpha = 0.0113$ derived from the parametrization in Fig.~\ref{params} (left), $\bar p_{ts} \approx 0.40$ GeV/c, $\bar p_{th0} \approx 1.3$ GeV/c and $\xi = 0.76$. The solid curve through points is the TCM described by Eqs.~(\ref{pampttcm}) and (\ref{xmodel})  with parameters  $\alpha = 0.0113$ and $\bar p_{th0} = 1.3$ GeV/c held fixed as for 5 TeV \pp\ collisions (assuming no jet modification).  Parameters $\bar \rho_{s0} \approx 3 \bar \rho_{sNSD} \approx 15$ and $m_0 \approx 0.10$  accurately describe the \pa\ \mmpt\ data. Exponent $n_1=5$ affects the TCM shape only near the transition point.

Figure~\ref{padata} (right) shows data in the left panel converted to $(n_{ch}' / n_s)\, \bar p_t' \approx \bar P_t / n_{s}$ by factor $\xi + x(n_s) \nu (n_s)$ as in Eq.~(\ref{nchppb}) (fifth line). The dashed line is the TCM for 5 TeV \pp\ collisions defined by Eq.~(\ref{niceeq}). The solid curve is the \ppb\ TCM defined by Eq.~(\ref{pampttcm})  (second line) derived from the solid curve in the left panel. Transforming the data from left to right panels requires an estimate of $n_s$ for the data to evaluate the required conversion factor $\xi + x(n_s) \nu(n_s)$. The map $n_s \rightarrow n_{ch}$ for the TCM from Eq.~(\ref{nchppb}) (second line) is inverted  via linear interpolation to provide the map $n_{ch} \rightarrow n_s$ for data. The accuracy of the \ppb\ TCM is apparent. There is little leeway in the two-parameter ($\bar \rho_{s0}, m_0$) \pa\ TCM once the \pp\ TCM is established as in Sec.~\ref{pptcmm}.

\subsection{p-Pb TCM conclusions}

\ppb\ \mmpt\ data appear to confirm that the soft yield $n_s$ or angular density $\bar \rho_s$ representing participant low-$x$ gluons is the basic parameter unifying \pp, \pa\ and \aa\ collision systems. Corresponding \pp\ spectrum soft component $S_{pp}(p_t,n_s)$ is universal, and its ensemble mean $\bar p_{ts} \approx 0.40$ GeV/c is equivalent to the same slope parameter $T \approx 145$ MeV for all systems, with minor variation of soft-component $\hat S_0(p_t)$ exponent $n$ vs collision energy~\cite{alicetomspec}.

\ppb\ \mmpt\ data also provide some understanding of the transition from isolated \nn\ collisions to the geometry of compound A-B systems, from noneikonal \pp\ to eikonal \aa\ Glauber model. Factorization of soft density $\bar \rho_s = \bar \rho_{sNN}(n_s) \, N_{part}(n_s)/2 $ and hard density $\bar \rho_h = \bar \rho_{hNN}(n_s) \, N_{bin}(n_s)$ distinguishes \nn\ internal structure ($\bar \rho_{sNN}$, $\bar \rho_{hNN}$) from A-B geometry ($N_{part}$, $N_{bin}$),
\bea \label{xnuu}
\frac{\bar \rho_h(n_s)}{\bar \rho_s} &=& \frac{ \bar \rho_{hNN}(n_s)}{\bar \rho_{sNN}(n_s)}\, \frac{N_{bin}(n_s)}{N_{part}(n_s)/2} \equiv x(n_s) \nu(n_s).~~
\eea
For the \pa\ TCM \pp\ noneikonal dijet production applied to \pa\ collisions averaged over \nn\ collisions and described by $\bar \rho_{hNN} = \alpha \bar \rho_{sNN}^2$ is a key assumption, and
\bea
x(n_s) &\approx&  \frac{\alpha \bar \rho_{s}}{N_{part}/2}
\eea
determines $N_{part}(n_s)$ given a model for $x(n_s)$. The \pa\ $x(n_s)$ model is the simplest extrapolation of the \pp\ $x(n_s) \approx \alpha \bar \rho_s$ linear trend possible: a continuing linear trend but with reduced slope beyond a transition point as in Eq.~(\ref{xmodel}). For \pa\ collisions the other Glauber parameters are immediately determined therefrom as in Sec.~\ref{patcm} above. The $x(n_s)$ and $\nu(n_s)$ models for the \pa\ TCM anticipate details for peripheral \aa\ collisions. 

Generally, as charge multiplicity (and hence $n_s$) increases for a given A-B system the product $x(n_s) \,\nu(n_s)$ measures key changes in hadron production. For \pp\ collisions $x(n_s) \approx \alpha \bar \rho_s$ carries essential information and $\nu \equiv 1$. For \aa\ collisions $x(n_s)$ is slowly varying (in the absence of jet modifications) and large $\nu(n_s)$ variation measures \aa\ geometry evolution within the Glauber model. As noted, the \pa\ case is intermediate with large excursions of $x(n_s)$ while $\nu(n_s)$ varies more slowly.

The \ppb\ TCM indicates that MB dijets dominate \pa\ collisions as they do \pp\ collisions because of the central role played by the noneikonal $\bar \rho_{hNN} \approx \alpha \bar \rho_{sNN}^2$ relation. For the data in Fig.~\ref{padata} the dijet yield for maximum value $\bar \rho_{sNN} = x(n_s) / \alpha \approx 22 \approx 4.4 \times \bar \rho_{sNSD}$ is 20 times that for NSD \pp\ collisions at the same energy. For 5 TeV \pp\ collisions the equivalent factor is 50. The result is large \mmpt\ variations due to MB dijets as in Fig.~\ref{padata}, described accurately by a simple \ppb\ TCM with two fixed parameters $(\bar \rho_{s0},m_0)$. In effect, \mmpt\ variation with \nch\ provides an indicator to establish factorization between \nn\ dijet production and the A-B Glauber description.

That result may be contrasted with a Glauber analysis of \ppb\ centrality in Ref.~\cite{aliceppbprod} based on the assumption that  ``...the multiplicity of charged particles at mid-rapidity [in \ppb] scales linearly with the total number of [nucleon] participants....'' The assumption is equivalent to the statement that all \nn\ collisions are the same, which is contradicted by \pp\ data. The Glauber analysis predicts rapid increase of $N_{part}(n_{ch})$ and $\nu(n_{ch})$ for lower \nch\ in contrast to the inferred TCM \ppb\ trend for $\nu(n_s)$ in Fig.~\ref{paforms} (right). In effect, the TCM product $x(n_s) \nu(n_s)$ that describes various \pa\ trends, including $\bar p_t(n_{ch})$, includes rapid increase of $x(n_s)$ ($\propto \bar \rho_s$) and slow increase of $\nu(n_s)$, whereas the Glauber analysis predicts the opposite. The proper choice depends in part on measured $dP /dn_{ch}$ for \pn\ (\pp) collisions, which is available from Ref.~\cite{alicemultpp}, compared to the \pa\ Glauber $dP/dN_{part}$.

\section{$\bf \bar p_t$ TCM for $\bf Pb$-$\bf Pb$ collisions} \label{pbpb}

Analysis of  \aa\ collisions requires a description of \aa\ geometry in terms of \nn\ binary collisions. \aa\ geometry is conventionally described by a Glauber Monte Carlo model based on the eikonal approximation. The number of participant nucleons N is represented by $N_{part}$, the number of \nn\ binary collisions by $N_{bin}$ and the number of binary collisions {\em per participant pair} by $\nu \equiv 2N_{bin} / N_{part} \approx (N_{part}/2)^{1/3} \in [1,6]$ (for A $\approx 200$). Glauber parameters may be defined in terms of fractional cross section $\sigma / \sigma_0$ via power-law relations as in Ref.~\cite{powerlaw}. 
The \aa\ Glauber model typically assumes that all \nn\ encounters within \aa\ collisions are equivalent, but in the present treatment the \aa\ description for more-peripheral collisions is modified to reflect lessons from \ppb\ data presented in the previous section.

\subsection{\pbpb\ $\bf n_{ch}$ TCM}

The TCM for total charge \nch\ integrated within angular acceptance $2\pi$ azimuth and $\Delta \eta$ pseudorapidity is given by Eqs.~(\ref{nchppb}).
In contrast to the \pa\ case $x(n_s) \equiv n_{hNN}(n_s) / n_{sNN}(n_s)$ can be estimated directly from measured \pbpb\ yield data vs centrality, with $\nu$ vs $n_{ch}$ determined by the Glauber model for \aa\ collisions following a two-step procedure as described below. 

\begin{figure}[h]
  \includegraphics[width=1.65in,height=1.6in]{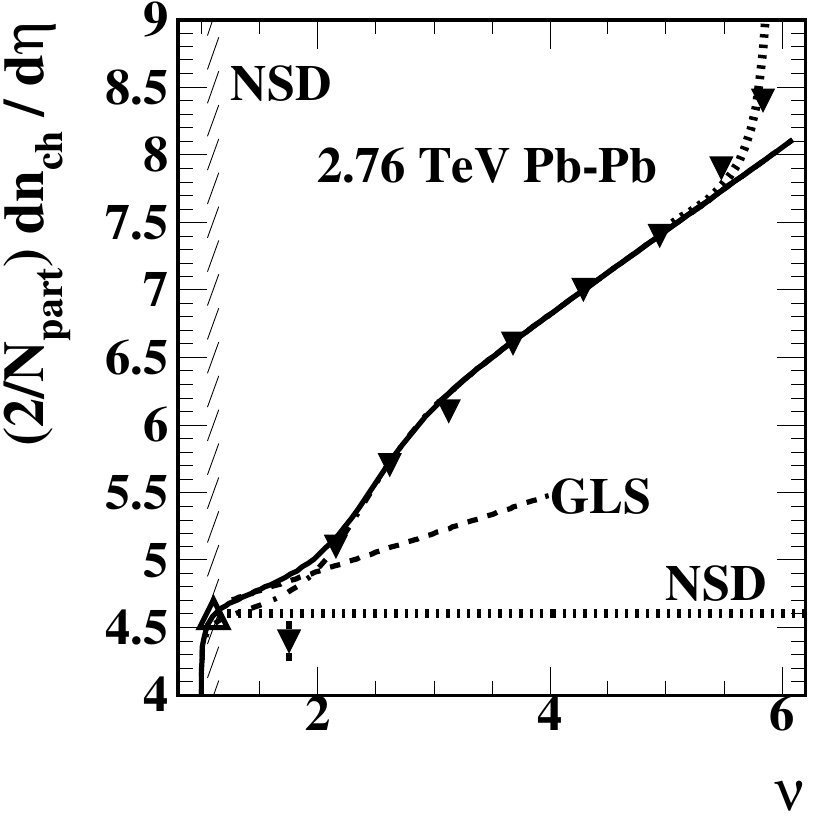}
  \includegraphics[width=1.65in,height=1.6in]{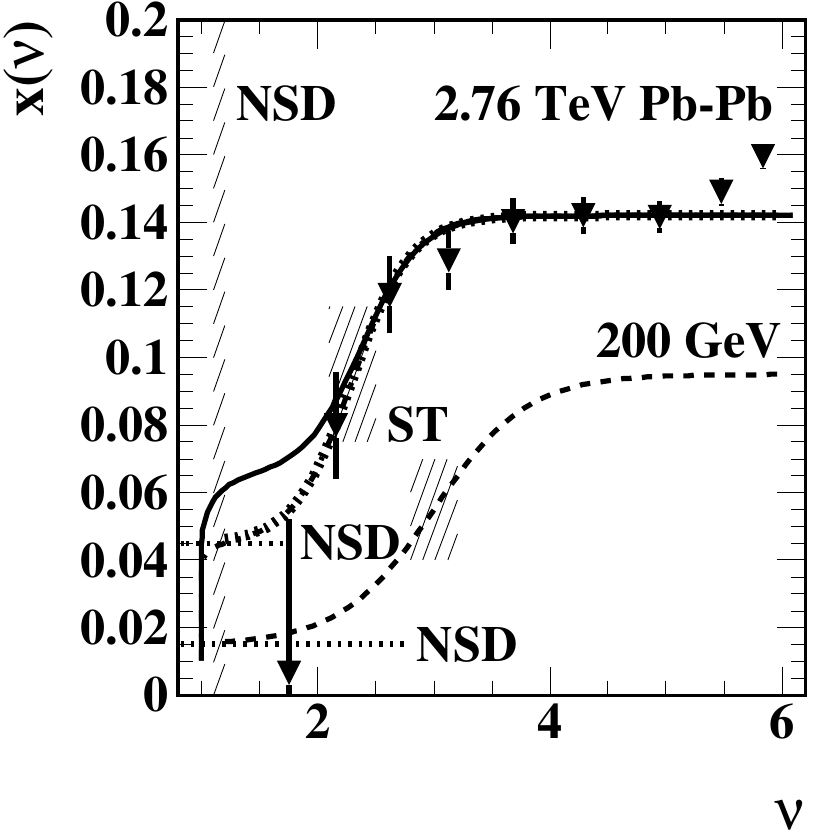}
  \caption{\label{nchaa}
  Left:
 Per-participant-pair hadron production vs centrality (binary collisions per participant pair $\nu$) for 2.76 TeV \pbpb\ collisions (inverted solid triangles~\cite{alicemultpbpb}) compared to the TCM trend of Eq.~(\ref{nchppb}) (third line, solid curve). The dash-dotted curve is explained in the text. The upright triangle is a 2.76 TeV NSD \pp\ reference. The GLS reference (dashed) is an extrapolation of the peripheral \nn\ TCM assuming no jet modification. The bold dotted curve at right indicates the effect of multiplicity fluctuations~\cite{tomphenix}.
  Right: Coefficient $x(\nu)$ (points) obtained by inverting  data from  the left panel  with Eq.~(\ref{nchppb}) (third line), and the $\tanh$ model function for $x(\nu)$ that generates the solid curve in the left panel (solid). The bold dotted curve represents Eq.~(\ref{xnu}). The dashed curve is the 200 GeV equivalent. The dash-dotted curve is explained in the text. The hatched bands indicate STs (e.g.\ as observed for 200 GeV  jet-related correlations~\cite{anomalous}).
   } 
  \end{figure}

Figure~\ref{nchaa} (left) shows per-participant-pair charge angular density vs mean participant path length $\nu$ for 2.76 TeV \pbpb\ collisions (inverted triangles)~\cite{alicemultpbpb}. 
The solid curve is the \nch\ TCM for \pbpb\ collisions from Eq.~(\ref{nchppb}) (third line) with $x(n_s)$ (derived below from data in this figure) and $\nu(n_s)$ described by the solid curves in Fig.~\ref{paglauber}. The dash-dotted curve is derived from the corresponding curve in  Fig.~\ref{paglauber} (right). 
The dashed line is a Glauber linear superposition (GLS) extrapolation from inelastic \nn\ (p-p) collisions with $x(n_s) \rightarrow x_{NSD}$ (assuming that jets are unmodified in \aa\ collisions). The upright triangle is an NSD \pp\ reference. The bold dotted curve shows the effect of multiplicity fluctuations for given \aa\ centrality appearing at the endpoint ($b \approx 0$) of the minimum-bias distribution $d\sigma/dn_{ch}$~\cite{tomphenix}, in this case corresponding to $|\eta|< 0.3$ as in Ref.~\cite{alicempt}. Fluctuation effects for this quantity are larger for smaller $\eta$ acceptance.

Figure~\ref{nchaa} (right) shows values of $x(\nu)$ (points) inferred from data in the left panel by inverting  Eq.~(\ref{nchppb}) (third line) assuming $\nu$ vs centrality defined by the \pbpb\ Glauber model and  $\bar \rho_{sNN} \approx \bar \rho_{sNSD} \approx 4.35 $~\cite{alicespec}. The bold dotted curve is a TCM model function describing the data,
\bea \label{xnu}
x(\nu) &=& 0.045 + 0.097 \left\{1+ \tanh[(\nu - 2.3)/0.5] \right\} /2.~~~~
\eea
with  {\em sharp transition} (ST) near $\nu = 2.3$ indicated by the upper hatched band.
The dashed curve is the $x(\nu)$ trend for 200 GeV \auau\ collisions with ST near $\nu = 3$~\cite{anomalous} (lower hatched band).
Note that the central-to-peripheral ratio of $x$ values for 200 GeV $\approx 6$ is twice as large as that for 2.76 TeV $\approx 3$ suggesting possibly {\em stronger} jet modification for lower collision energies.

The dash-dotted curve (just visible) is the bold dotted curve combined with the \pp\ trend $x(n_s) \approx \alpha \bar \rho_s$ as in Eq.~(\ref{xpbpb}) below. The solid curve corresponds to the \pp\ trend extended beyond $\bar \rho_{sNSD}$ to some transition point $\bar \rho_{s0}$ (as for \ppb\ collisions). These hadron production data alone cannot determine $\bar \rho_{s0}$; an additional constraint is identified below.  The solid and dash-dotted curves describing $x(\nu)$ in this panel are obtained by combining the corresponding curves for $x(n_s)$ and $\nu(n_s)$ derived in the next subsection and shown in Fig.~\ref{paglauber}.  The NSD limiting values for $\nu \approx 1$ (dotted lines) are determined by $x_{NSD} \approx \alpha \bar \rho_{sNSD}$, where $\alpha = 0.0103$ from Fig.~\ref{params} (left) and $\bar \rho_{sNSD} = 4.35$ from Ref.~\cite{alicespec} give $x_{NSD} \approx 0.045$ for 2.76 TeV. The corresponding numbers for 200 GeV \pp\ collisions are $x_{NSD} \approx 0.006 \times 2.5 = 0.015$.

\subsection{Pb-Pb Glauber-model modifications from p-Pb}

\begin{figure}[h]
  \includegraphics[width=1.65in]{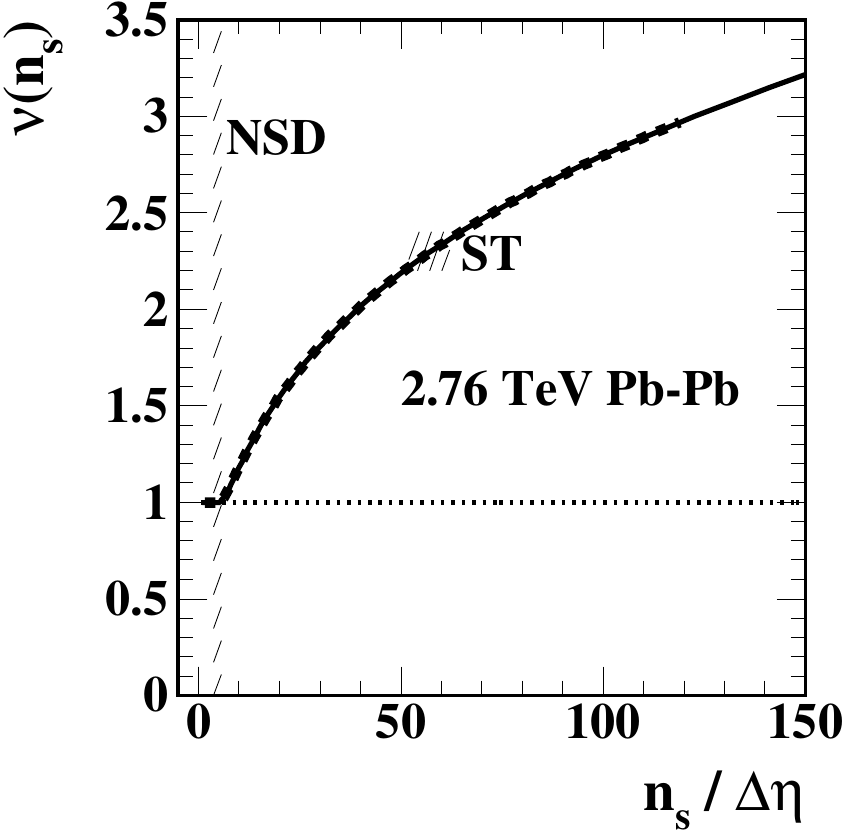}
   \includegraphics[width=1.65in]{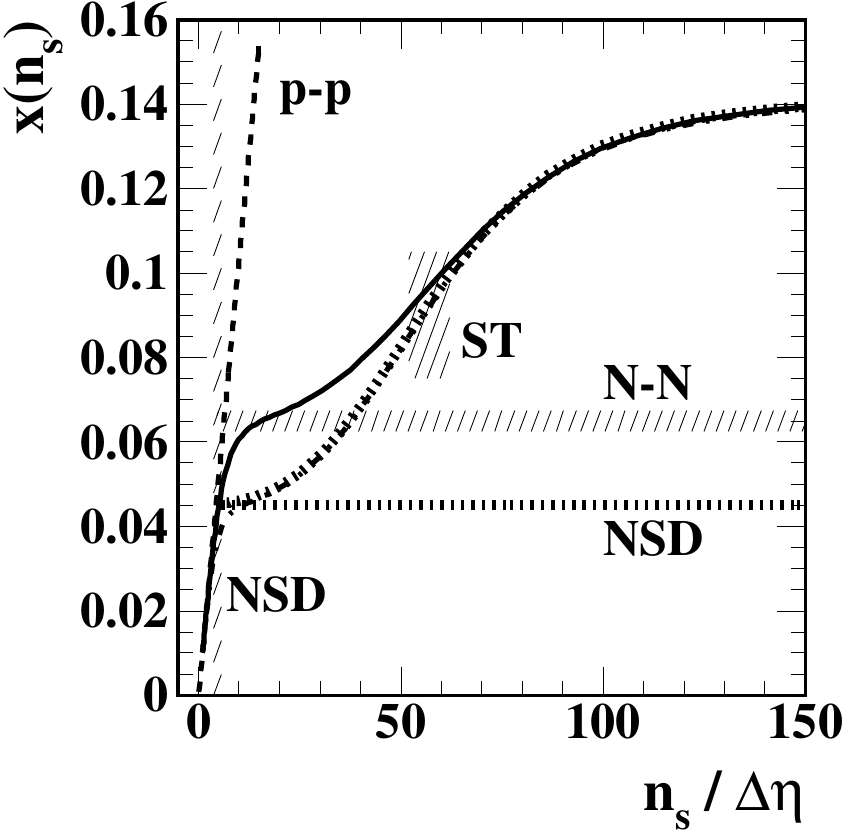}
  \caption{\label{paglauber}
  Left: Mean participant pathlength $\nu(n_s) \approx (\bar \rho_s / \bar \rho_{sNSD})^{1/3}$ (solid curve) for 2.76 TeV \pbpb\ collisions consistent with the Glauber model based on an eikonal approximation~\cite{powerlaw}.
  Right: TCM parameter $x(n_s)$ derived from a hadron production trend (bold dotted curve) equivalent to the same curve on $\nu$ in Fig.~\ref{nchaa} (right). The dash-dotted curve (just visible) is the dotted curve combined with a \pp\ noneikonal trend as in Eq.~(\ref{xpbpb}). The solid curve is the same but with a transition point $\bar \rho_{s0} > \bar \rho_{sNSD}$.
   }  
 \end{figure}

Figure~\ref{paglauber} is modified from Fig.~\ref{paforms} to combine Glauber \aa\ trends as in Eq.~(\ref{xnu}) with \pa\ trends from Sec.~\ref{ppb}. 

Figure~\ref{paglauber} (left) shows mean participant pathlength $\nu(n_s)$ (solid curve), identical to the dash-dotted curve in Fig.~\ref{paforms} (right), defined by
\bea \label{nupbpb}
\nu(n_s) &=&  \left\{1 + [h(n_s)]^{n_3}\right\}^{1/n_3}
\eea 
with  $h(n_s) = (\bar \rho_s/ \bar \rho_{sNSD})^{1/3}$, $\bar \rho_{sNSD} = 4.35$ and $n_3 = 5$.
The $h(n_s)$ trend is determined by the naive \aa\ Glauber model.
Whereas the solid curve represents the TCM with map $n_s \rightarrow n_{ch}$ the bold dashed curve is defined on data \nch\ values with map $n_{ch} \rightarrow n_s$, by interpolation via Eqs.~(\ref{nchppb}), to demonstrate analysis consistency. 

Figure~\ref{paglauber} (right) shows the TCM hard/soft parameter  $x(n_s)$ (solid curve) derived from \ppb\ and \pbpb\ data. For a naive \pbpb\ Glauber model and \nn\ linear superposition (GLS)  $x(n_s)$ would retain the NSD value $\alpha \bar \rho_{sNSD} = 0.0103 \times 4.35 \approx 0.045$ (dotted line).  The hadron-yield centrality trend in Fig.~\ref{nchaa} indicates that because of modifications to jet formation (``jet quenching'') in \aa\ collisions $x(n_s)$ increases substantially [bold dotted curve defined by Eq.~(\ref{xnu})] indicating a rapid change in jet properties~\cite{anomalous} and thereafter remains constant for more-central \pbpb\ collisions. 
The lessons from \ppb\ collisions for {\em peripheral} \pbpb\ collisions are as follows: (a) For $\bar \rho_s < \bar \rho_{sNSD}$ the \pp\ trend $x(n_s) \approx \alpha \bar \rho_s$ (dashed line) applies. Combination with the bold dotted curve leads to the dash-dotted curve. (b) For $\bar \rho_s > \bar \rho_{sNSD}$ $x(n_s) \approx \alpha \bar \rho_s$ may continue to some transition point $\bar \rho_{s0}$ (corresponding to the \nn\ hatched band).  To accommodate point (b) the dash-dotted curve is transformed to the solid curve.

The full TCM expression for $x(n_s)$ in Fig.~\ref{paglauber} is
\bea \label{xpbpb}
x(n_s) &=& \frac{1}{\left\{[1/\alpha \bar \rho_s]^{n_2} + [1/g(n_s)]^{n_2}\right\}^{1/n_2}},
\eea
where $\alpha \bar \rho_s$ is the \pp\ trend as in Eq.~(\ref{xmodel}) but $n_2 = 3$ and
\bea
g(n_s) &=& x_{pp} +
\\ \nonumber
&& (0.142-x_{pp})\{\tanh[(\nu(n_s)-2.38)/0.5]+1\}/2
\eea
with $\nu(n_s)$ defined in the left panel.  $x_{pp} \approx 0.065$ corresponds to transition point $\bar \rho_{s0}$ = 6.3 (solid curve, \nn\ hatched band). If $\bar \rho_{s0} \rightarrow \bar \rho_{sNSD} = 4.35$ then $x_{pp} \rightarrow 0.045$ (dash-dotted curve, NSD dotted line). Parameter values are inferred from \pbpb\  data.  The curves in Fig.~\ref{paglauber} (right) on $\bar \rho_s$ and Fig.~\ref{nchaa} (right) on $\nu(n_s)$ are equivalent.


 
\subsection{Formulating a \pbpb\ $\bf \bar P_t$ TCM}

Referring to Eq.~(\ref{xeqn}) the TCM for the ensemble mean of extensive variable  $Z \rightarrow P_t$ in \aa\ collisions is
\bea \label{nchaaa}
\frac{2}{N_{part}} \bar P_t(n_s) &=& \bar P_{tsNN}(n_s) + \nu(n_s)\, \bar P_{thNN}(n_s)
\\ \nonumber &=& n_{sNN}(n_s)[ \bar p_{ts} + x(n_s) \, \nu(n_s) \bar p_{thNN}(n_s)],
\eea
where $\bar p_{ts} \approx 0.40$ GeV/c~\cite{ppprd} appears to be a common feature of all high-energy nuclear collisions~\cite{jetspec2}. The nominal value of $n_{sNN}(n_s)$ is moot since it only appears here in the product $n_s =(N_{part}/2) n_{sNN}(n_s)$. Both $x(n_s)$ and $\bar p_{thNN}(n_s)$ are expected to evolve with \aa\ centrality per jet modification, and the details for 2.76 TeV \pbpb\ collisions are a goal of the present study (see Sec.~\ref{pbpbdetails}). 

The \pbpb\ $\bar p_t'$ data from Ref.~\cite{alicempt} have the same general TCM description as for \ppb\ data
\bea \label{ptprime}
\bar p_t' \equiv \frac{\bar P_t'} {n_{ch}'} &\approx & \frac{\bar P_{tsNN}' + \nu(n_s) \bar P_{thNN}(n_s)}{n_{sNN}'(n_s) + \nu(n_s) n_{hNN}(n_s)}~~~
\\ \nonumber
&\approx &  \frac{\bar p_{ts} + x(n_s)\, \nu(n_s) \bar p_{thNN}(n_s)}{\xi + x(n_s) \nu(n_s)},
\eea
where primes denote the effects of uncorrected data integrated over a \pt\ interval with lower acceptance limit $p_{t,cut} \approx 0.15$ GeV/c. 
The details of \pt\ production can be isolated by forming the product $(n_{ch}' / n_{s}) \, \bar p_t'$ to obtain
\bea \label{proddd}
\frac{n_{ch}' }{n_{s}}  \bar p_t'(n_s) &\approx& \frac{\bar P_t}{n_{s}} \approx \bar p_{ts} + x(n_s)   \,  \nu(n_s)\, \bar p_{thNN}(n_s).~
\eea
From $\nu(n_s)$ and $x(n_s)$ as in Fig.~\ref{paglauber} \nch\ values can be derived from any set of $n_s$ values via Eq.~(\ref{nchppb}) (third line) for TCM quantities. Given that relation the map $n_{ch} \rightarrow n_s$ for data values of \nch\ is established by interpolation.

\subsection{\pbpb\ $\bf \bar p_t'$ data and models} \label{pbpbtrends}

\begin{figure}[h]
  \includegraphics[width=1.65in,height=1.6in]{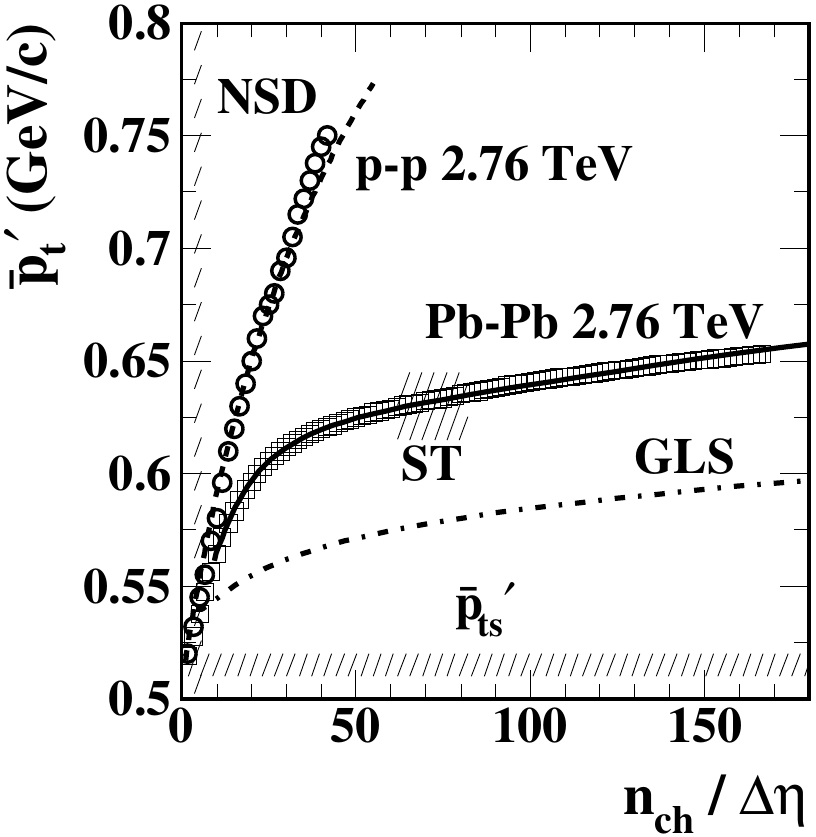}
  \includegraphics[width=1.65in,height=1.6in]{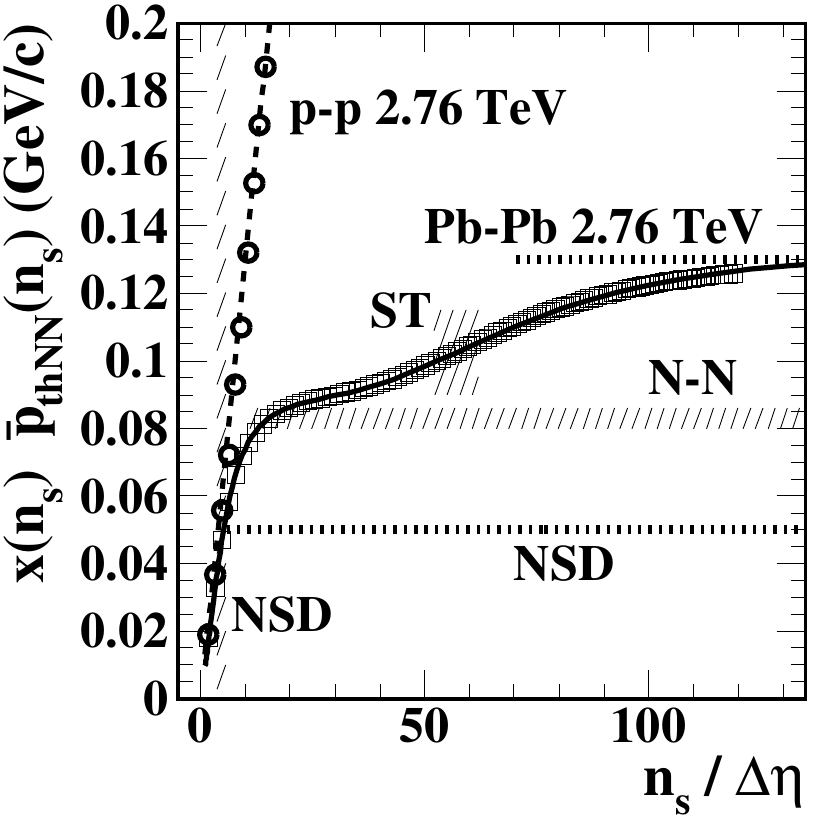}
  \caption{\label{mptxx}
  Left: $\bar p_t'$ data for 2.76 TeV \pbpb\ collisions from Ref.~\cite{alicempt} (open squares). The open circles and dashed curve are \pp\ data and TCM from Sec.~\ref{pptcmm}. The solid curve is the \pbpb\ TCM. The dash-dotted curve is a GLS reference assuming all \nn\ collisions are equivalent to NSD \pp\ collisions and scale according to the Glauber model of \pbpb\ collisions following the eikonal approximation. The lower hatched band estimates $\bar p_{ts}' \approx \bar p_{ts} / \xi$. The upper hatched band indicates the location on \nch\ of a ST appearing in Fig.~\ref{nchaa} (right) on $\nu$.
  Right: The product $x(n_s)\,\bar p_{thNN}(n_s) \approx [\bar P_{t}(n_s) / n_{s} - \bar p_{ts}] / \nu(n_s)$ extracted from  data and TCMs in the left panel according to Eq.~(\ref{proddd}), with $x_{NSD} \,\bar p_{thNSD}  \approx 0.045 \times 1.1 \approx 0.05$ GeV/c (dotted) and $x_{NN} \,\bar p_{thNN}  \approx 0.065 \times 1.28 = 0.083$ GeV/c (hatched). The ST is apparent in this format but not in the left panel.
   }  
  \end{figure}

\begin{figure}[h]
  \includegraphics[width=1.68in]{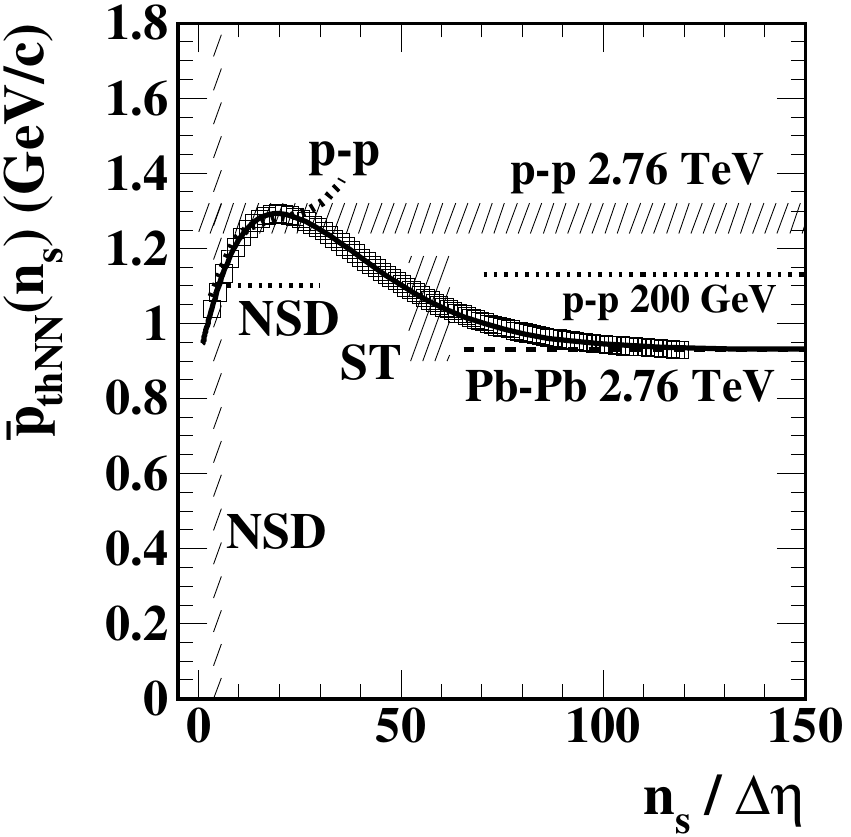}
  \includegraphics[width=1.68in]{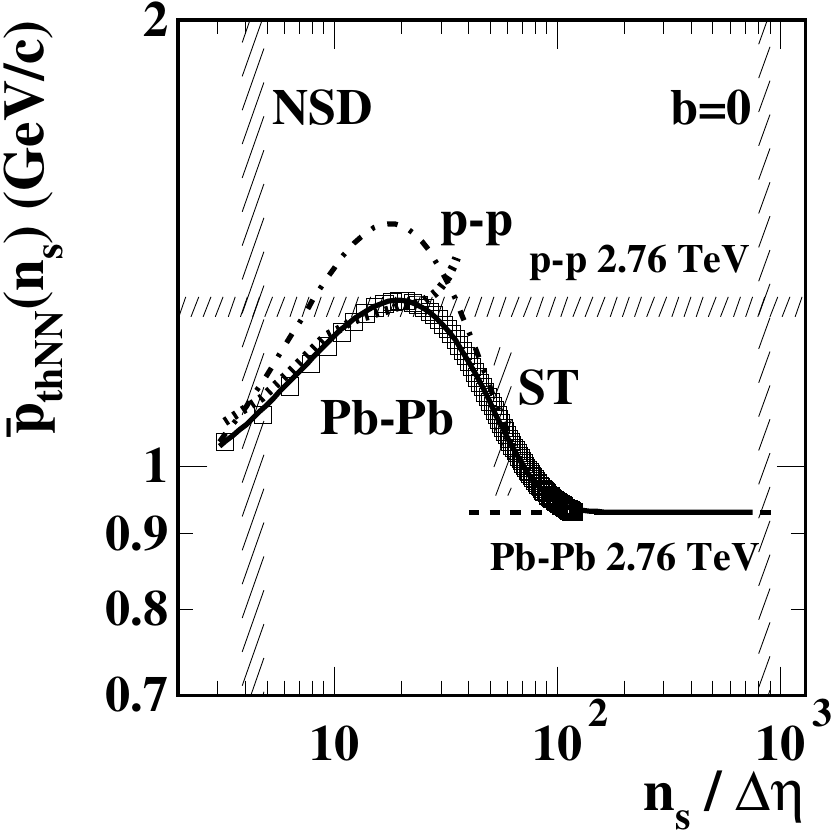}
  \caption{\label{mptyy}
  Left: \mmpt\ hard component $\bar p_{thNN}(n_s)$ (points) inferred from 2.76 TeV \pbpb\ $\bar p_t'$ data by inverting Eq.~(\ref{ptprime}) with TCM parameters $x(n_s)$ and $\nu(n_s)$ defined in Fig.~\ref{paglauber}. The solid curve is defined by Eq.~(\ref{pthnn}).
  Right:  Data and curves in the left panel in a log-log format. The dash-dotted curve indicates where the data points would lie if the dash-dotted $x(n_s)$ curve in Fig.~\ref{paglauber} (right) were used to transform the $\bar p_t'$ data. The bold dotted curve in both panels is the 2.76 TeV \pp\ data trend from Fig.~\ref{alice5} (right) used to define transition point $\bar \rho_{s0} = 6.3$.
   }  
  \end{figure}

Figure~\ref{mptxx} (left) shows $\bar p_t'$ data from Ref.~\cite{alicempt} for 2.76 TeV \pbpb\ collisions (open squares) and \pp\ collisions (open circles) vs corrected $n_{ch}$ in the form $\bar \rho_0 = n_{ch} / \Delta \eta$, where the increase of $\bar p_t'$ in \aa\ collisions was attributed in that paper to radial flow increasing with collision centrality: ``most particles are part of a locally thermalized medium exhibiting collective, hydrodynamic-type, behavior.'' The limiting case for small \nch\ is $\bar p_{ts}' = \bar p_{ts}/ \xi \approx 0.515$ GeV/c. The \pp\ dashed curve is defined by Eq.~(\ref{ppmpttcm}).  The \pbpb\ solid curve is the TCM defined by Eq.~(\ref{ptprime}) with $x(n_s)$ and $\nu(n_s)$ defined by  the solid curves in Fig.~\ref{paglauber} and $\bar p_{thNN}(n_s)$ defined below by the solid curves in Fig.~\ref{mptyy}. There is no obvious manifestation in this format of the ST near $n_{ch} \approx 70$ (hatched band) observed in Fig.~\ref{nchaa} (right) near $\nu = 2.3$. The dash-dotted curve is a GLS reference defined by Eq.~(\ref{ptprime})  with $x = 0.045$ and $\bar p_{thNN} = 1.1$ GeV/c held fixed at NSD values and no jet modification. The result is comparable to the \pbpb\ HIJING trend in Fig.~3 of Ref.~\cite{alicempt}.

Figure~\ref{mptxx} (right) shows data from the left panel transformed (points) according to the expression
\bea
x(n_s) \, \bar p_{thNN}(n_s) &\approx&  \frac{1}{\nu(n_s) }  \left[\frac{n_{ch}' }{n_{s}}  \bar p_t'(n_s) - \bar p_{ts}\right]
\eea
with $\nu(n_s)$ defined by the solid curve in Fig.~\ref{paglauber} (left) to isolate a product relevant to \pt\ production via MB dijets in a relatively model-independent way. In this differential format the ST near $\bar \rho_s \approx 58$ for 2.76 TeV \pbpb\ collisions is evident.
The same product for NSD \nn\ collisions is
\bea
x(n_s) \, \bar p_{thNN}(n_s) &\rightarrow&  \alpha \bar \rho_{sNSD}\, \bar p_{thNSD} 
\\ \nonumber
&& \hspace{-.5in} \approx~ 0.0103   \times 4.35 \times 1.1 \approx 0.05~ \text{GeV/c}
\eea
indicated by the dotted line. The hatched band represents a \nn\ reference with $\bar \rho_s \rightarrow \bar \rho_{s0} \approx 6.3$ and $ \bar p_{thNN} \rightarrow \bar p_{th0} \approx 1.28$ GeV/c; the product above then goes to 0.083 GeV/c.  Given a TCM expression for $x(n_s)$, data in the right panel can be used to obtain $\bar p_{thNN}(n_s)$. It is interesting that $x(n_s) \, \bar p_{thNN}(n_s) \approx \bar P_{thNN}(n_s) / n_s$, the total \mmpt\ hard component per \nn\ collision and per soft hadron measuring the dijet contribution to \mmpt, {\em increases substantially} with centrality in \pbpb\ collisions.

Figure~\ref{mptyy} (left) shows the \pbpb\ data in Fig.~\ref{mptxx} (right) divided by $x(n_s)$ defined by the solid curve in Fig.~\ref{paglauber} (right). The bold dotted curve (just visible) represents the 2.76 TeV \pp\ $\bar p_{th}(n_s)$ data from Fig.~\ref{alice5} (right). The horizontal hatched band represents the $p_{th0} \approx 1.28$ GeV/c \pp\ value  from Fig.~\ref{enrat3} (right). For more-central collisions the data approach asymptotically a lower limit 0.93 GeV/c (dashed line). The 200 GeV value $\bar p_{t0} \approx 1.13$ GeV/c is included for comparison (right dotted line). 
 
Figure~\ref{mptyy} (right) is the left panel replotted in a log-log format covering the entire centrality range of \pbpb\ collisions up to $b = 0$  (most central) and providing a more-detailed view of the peripheral region. The most-central asymptotic limit 0.93 GeV/c (dashed line) is about 73\% of $\bar p_{th0} \approx 1.28$ GeV/c. The dash-dotted curve is the trend the data would follow if transformed by the dash-dotted curve for $x(n_s)$ in Fig.~\ref{paglauber} (right) without extension of the \pp\  trend to some $\bar \rho_{s0} > \bar \rho_{sNSD}$. The inferred $\bar p_{thNN}(n_s)$ trend would then disagree with the measured \pp\ trend (bold dotted curve). The assumed criterion that the $\bar p_{thNN}(n_s)$ trend in peripheral \pbpb\ collisions should match the $\bar p_{th}(n_s)$ trend in \pp\ collisions determines the value $\bar \rho_{s0} = 6.3$ marking a transition from isolated \nn\ collisions to the \aa\ Glauber model that defines the $x(n_s)$ solid curve in Fig.~\ref{paglauber} (right). A parametrization of the $p_{thNN}(n_s)$ data (solid curve) is given by
\bea \label{pthnn}
\bar p_{thNN}(n_s) &=&  \{1+\tanh[(\bar \rho_s+3.5)/14]\}/2 \times
\\ \nonumber
&& \hspace{-.5in} 1.53\left(1-0.39\{1+\tanh[(\bar \rho_s-34)/36]\}/2\right).
\eea 
The expressions for $\nu(n_s)$ in Eq.~(\ref{nupbpb}),
$x(n_s)$ in Eq.~(\ref{xpbpb})
 and  $\bar p_{thNN}(n_s)$ in Eq.~(\ref{pthnn}) combined in universal Eq.~(\ref{pampttcm})  constitute the TCM for \mmpt\ data from 2.76 TeV \pbpb\ collisions that defines all solid curves in this section.

\subsection{Pb-Pb TCM conclusions}

For \aa\ collisions the trends $N_{bin} \approx (N_{part}/2)^{4/3}$ and  $\nu \approx (N_{part}/2)^{1/3}$ are manifestations of the naive eikonal-based Glauber model that serves as the preferred \aa\ centrality estimator assuming that all \nn\ encounters are the same [i.e.\ $\rho_{sNN}(n_s) \approx \bar \rho_{sNSD}$]. But that assumption and certain details for peripheral collisions should be revisited  based on \pp\ and \ppb\ experience described above. The transition from noneikonal \pp\ to eikonal \aa\ as limiting cases is important for any A-B system.

\aa\ collisions manifest two transitions: (a) from \pp\ (or \nn) to A-B geometry still retaining  ``in-vacuum'' jet formation and (b) onset of jet modification at a ST not evident in \pa\ collisions. Two analysis questions thus emerge: (a) For more-peripheral A-B collisions where is the transition point $\bar \rho_{so}$ between \pp\ and alternative trends? (b) For more-central collisions where (and how) do jet properties transition from ``in-vacuum'' jet production to the jet modification observed in \aa\ near ST?

For case (a) \ppb\ data establish an essential intermediary for the \pbpb\ model, but $\bar \rho_{s0}$ appears substantially lower: $\approx 1.45\,\bar \rho_{sNSD}$ in \pbpb\ vs $3 \,\bar \rho_{sNSD}$ in \ppb. For case (b) hadron production trends as in Fig.~\ref{nchaa} are determining. Thus, \ppb\ results determine the \pbpb\ model for $\bar \rho_s < 30$, and measured hadron production plus naive Glauber determine the model for  $\bar \rho_s > 30$. Those statements apply to model elements $x(n_s)$ and $\nu(n_s)$. It then remains to characterize hard component $\bar p_{thNN}(n_s)$ derived from \pbpb\ \mmpt\ data, as in Fig.~\ref{mptyy} and Eq.~(\ref{pthnn}).

The resulting TCM hard-component yield (jet fragments) per \nn\ collision increases with centrality in \pbpb\ collisions (relative to NSD) by a factor three (Fig.~\ref{nchaa}) whereas the \mmpt\ hard component is only reduced by factor 0.85 (Fig.~\ref{mptyy}). Total-\pt\ hard component per soft hadron (low-$x$ gluon) $\bar P_{thNN}/n_s$ thus increases by factor 2.5 from peripheral (NSD \pp) to central \aa\ (Fig.~\ref{mptxx}), a large increase in specific (\nn) momentum transport from longitudinal to transverse phase space attributable to large-angle parton scattering and MB jet formation.

\section{Systematic uncertainties} \label{syserr}

The \mmpt\ TCM presented in this study includes quantities $\bar p_{ts}$, $\xi$, $x(n_s)$, $\nu(n_s)$ and $\bar p_{thNN}(n_s)$. This section reviews uncertainties in those quantities for each collision system.

\subsection{Uncertainties for particle data}

Statistical uncertainties are not relevant for LHC data, as evident for instance in Figs.~\ref{alice5a}, \ref{padata} and \ref{mptxx}. Principal uncertainties are then systematic. Since $\bar p_t$ is a ratio most systematic effects (e.g.\ tracking errors) cancel. The main issue is a vertical offset to $\bar p_t'$ data due to the {\em effective} lower-\pt\ cutoff which is not reported by experiments. That uncertainty is addressed in the next subsection.

\subsection{Uncertainties for p-p TCM}

The \pp\ spectrum TCM in Sec.~\ref{pptcmm} is required by data as demonstrated in Refs.~\cite{ppprd,ppquad}. Alternative spectrum models might be tuned to describe a particular spectrum but cannot describe the \nch\ dependence of spectrum structure~\cite{jetspec2,alicetomspec}. The same comment then applies to the statistic \mmpt\ derived implicitly from \pt\ spectra.

Given the approximately linear TCM demonstrated in Fig.~\ref{alice5} (left) the basic issues are uncertainties in the TCM intercept and slope. For uncorrected data in Fig.~\ref{alice5a} (left) the observed $n_{ch} \rightarrow 0$ intercept is $\bar p_{ts}' = \bar p_{ts} / \xi \approx 0.515$ GeV/c. A 5\% uncertainty in $\bar p_{ts} = 0.40 \pm 0.02$ GeV/c determines the uncertainty in $\xi$ which corresponds to a 10\% uncertainty in $p_{t,cut}$ or $0.17 \pm 0.02$ GeV/c as expected for TCM tracking inefficiency (Fig.~4, right).
Given the lower limit $p_{t,cut} = 0.15$ GeV/c the upper limit on $\xi$ is 0.83 as in Fig.~4 (right). Given the observed $\bar p_{ts}' \approx 0.515$ GeV/c the upper limit on $\bar p_{ts}$ is then 0.83 times 0.515 = 0.425 GeV/c. Reported in \cite{ppprd} is 0.385 GeV/c. The adopted estimate for this analysis as noted above is $\bar p_{ts} = 0.40 \pm 0.02$ GeV/c corresponding to $\xi = 0.77 \pm 0.04$.

The observed \pp\ TCM slope is the product of factors $\alpha(\sqrt{s}) \bar p_{th0}$. $\bar p_{th0}$ is predicted by measure \pt\ spectra with few-percent accuracy as in Fig.~\ref{enrat3}. Values of $\alpha(\sqrt{s})$ then inferred from \pp\ \mmpt\ data as in Fig.~\ref{params} (left) are consistent with previous spectrum analysis, and the energy dependence is consistent with jet measurements.

A novel feature of \pp\ collisions is the noneikonal trend for dijet production $\bar \rho_h \propto \bar \rho_s^2$ indicating that {\em each} participant parton (gluon) in one proton may interact with {\em any} participant in the partner proton, hence the quadratic dependence. In the eikonal approximation the expected trend would be approximately $\bar \rho_h \propto \bar \rho_s^{4/3}$. The noneikonal trend is manifested as the linear $\bar P_t / n_s$ trend in Fig.~\ref{alice5} (left) over an \nch\ interval corresponding to {\em 100-fold increase} in dijet production. The same trend is followed precisely for spectra~\cite{alicetomspec} and angular correlations~\cite{ppquad}.

A second feature is the \nch\ dependence of $\bar p_{th}(n_s)$ as in Fig.~\ref{alice5} (right) which might be interpreted as breakdown of noneikonal linearity (constant slope) of the $\bar P_t / n_s$ trend. However, as demonstrated in Figs.~\ref{300b} and \ref{300a} (right) the \nch\ dependence corresponds precisely to measured variation of the spectrum hard component attributed to bias of the underlying jet spectrum as measured by Figs.~\ref{newparams1} and \ref{newparams2}~\cite{alicetomspec}. The small deviations between data points and solid curves in Figs.~\ref{300b} and \ref{300a} indicate the overall self-consistency of the \pp\ TCM at the few-percent level.

\subsection{Uncertainties for p-Pb TCM}

For \ppb\ collisions the $x(n_s)$ model completely defines the TCM since all other quantities can be derived therefrom. The \ppb\ \mmpt\ trend below transition point $\bar \rho_{s0}$ follows the \pp\ trend with $x(n_s) \approx \alpha \bar \rho_s$ and inherits the \pp\ uncertainties  described above. Above the transition point the $x(n_s)$ model is a conjecture based on the simplest alternative: continued linear increase but with reduced slope. The TCM then has two adjustable parameters $(\bar \rho_{s0},m_0)$ with which to accommodate \mmpt\ data and does so at the percent level as demonstrated in Fig.~\ref{padata}.

The observed  $\bar p_{ts}' \approx 0.525 \pm 0.01$ GeV/c for \ppb\ data in Fig.~\ref{padata} (left) with $\xi \approx 0.76$ is consistent with the $\bar p_{ts}' \approx 0.515 \pm 0.01$ GeV/c for various \pp\ systems in Fig.~\ref{alice5a} (left) (corresponding to $\xi \approx 0.78$ and $p_{t,cut} \approx 0.17$ GeV/c in Fig.~4, right) within few-percent uncertainties.

The route from $x(n_s)$ to other TCM elements involves three basic assumptions: (a) The factorization in Eq.~(\ref{pafactor}) reflects a collision model wherein imposition of an \nch\ condition on \pa\ events leads to a compromise between higher individual \nn\ multiplicity and more \nn\ binary collisions, depending in turn on probability distributions on \nch\ for individual \nn\ collisions and on $N_{part}$ within \pa\ collisions. (b) The expression $\bar \rho_{hNN} \approx \alpha \bar \rho_{sNN}^2$ derived from isolated \pp\ collisions is still valid for \nn\ collisions within \pa\ collisions. (c) The inferred parameters $N_{part}$, $N_{bin}$ and $\nu$ based on the first two assumptions are meaningful in terms of \pa\ collision geometry in the sense of the Glauber model of composite collisions.

\subsection{Uncertainties for Pb-Pb TCM}

For \aa\ collisions TCM functions $\nu(n_s)$ and $x(n_s)$ can be derived independently. 
Mean participant pathlength $\nu$  is defined by geometry parameters $N_{part}$ and $N_{bin}$ which can be inferred for more-central \aa\ collisions from a naive Glauber model based on the eikonal approximation. The trend $\nu \approx (N_{part}/2)^{1/3}$ is observed~\cite{powerlaw}. $N_{part}$ is in turn related to parameter $n_s$ by factorization $\bar \rho_s = [N_{part}(n_s)/2]\, \bar \rho_{sNN}(n_s)$. 
Conventionally, $\bar \rho_{sNN}$ is assumed to have a fixed value within the \aa\ Glauber model corresponding to inelastic or NSD \pp\ collisions. For \pbpb\ \mmpt\ data $\bar \rho_{sNN} \approx \bar \rho_{sNSD}$ is assumed and $\bar \rho_s$ is mapped to \nch\ via Eqs.~(\ref{nchppb}) given $x(n_s)$ and $\nu(n_s)$. However, when applied to asymmetric systems such as \pa~\cite{aliceppbprod} that assumption is incorrect (see Sec.~\ref{ppb}).

$x(n_s)$ can be determined for more-central \aa\ collisions (at and above the ST) from the measured centrality trend of hadron production as in Fig.~14 (right) with uncertainty depending on the accuracy of such data. However, the centrality of more-peripheral collisions (below the ST) is poorly determined and information about $x(n_s)$ must be obtained elsewhere, e.g.\ \pp\ and \pa\ data. Specifically, $x(n_s)$ below the ST is assumed to have a trend similar to that for \ppb\ data but with possibly different transition point $\bar \rho_{s0}$ which must then be derived from \pbpb\ data. That determination in turn depends on \pp\ $\bar p_{th}(n_s)$ data as in Figs.~\ref{alice5} (right) and \ref{mptyy} (right) such that $\bar \rho_{s0} \approx 6.3$ is determined to a few percent.

 In Fig.~\ref{mptxx} (left) the observed $\bar p_{ts}' \approx 0.515$ GeV/c for \pbpb\ data agrees with the \pp\ and \ppb\ values noted above within uncertainties, indicating consistent detector performance and tracking analysis parameters. As with \ppb\ data the \pbpb\ data are congruent with \pp\ data for  \nch\ near and below the NSD value.

Given the conjecture $N_{part}(n_{ch}) \approx \bar \rho_{s} / \bar \rho_{sNN}(n_s)$ consistent with Eq.~(\ref{pafactor}) a principal uncertainty in  Fig.~\ref{mptxx} (right) arises from the model for $\nu(n_s) \approx (N_{part}/2)^{1/3}$, which depends on applicability of the naive Glauber model to more-central \aa\ collisions, and the specific value of $\bar \rho_{sNN}(n_s)$ used to define $\nu(n_s)$ in Eq.~(\ref{nupbpb}).  The choice $\bar \rho_{sNN}(n_s) \approx \bar \rho_{sNSD}$ is imposed by Fig.~\ref{nchaa} (left) where up to $\nu = 2$ the trend for $(2/N_{part}) \bar \rho_0 $ in Eq.~(\ref{nchppb}) (third line) is consistent with $\rho_{sNN}(n_s) < 1.1\, \bar \rho_{sNSD}$ since $x(n_s) \nu(n_s) \approx 0.05$ - 0.1 in that interval.  And that choice leads to $\nu(n_s)$ in Fig.~\ref{paglauber} (left) consistent with Glauber Monte Carlo calculations to a few percent. 

Given the assignment $\bar \rho_{sNN} \rightarrow  \bar \rho_{sNSD}$ the 50\% increase in $x(n_s)$ over its NSD value cannot then rely on the linear relation $x(n_s) \approx \alpha \bar \rho_s$ continuing past the NSD value as in \pa\ collisions. It must arise from a different mechanism in peripheral \pbpb\ introducing significant uncertainty for the TCM in peripheral \aa\ collisions that might be resolved by more-detailed peripheral data. It is interesting that the peripheral \pbpb\ \mmpt\ data trends (open squares) in Fig.~\ref{mptxx} follow the observed \pp\ trends (open circles) and the \pp\ TCM within small data uncertainties.

For central \pbpb\ $N_{part}/2 \approx 200$ to a few percent and $\nu \approx 5.9$ as a result. Given the $b = 0$ value $(2/N_{part}) \bar \rho_0 \approx 8$ in Fig.~\ref{nchaa} (left) the corresponding saturation value for $x(n_s)$ in the right panel is $(8/4.35 - 1)/ 5.9 = 0.142$ per Eq.~(\ref{nchppb}) (second line) also determined to a few percent.

 In Fig.~\ref{mptyy} the saturation (central) value for $\bar p_{thNN}(n_s)$ is obtained from Figs.~\ref{nchaa} (right) and \ref{mptxx} (right) as 0.132 / 0.142 = 0.93, one of the main results of the \pbpb\ analysis. The ST is evident at $\bar \rho_{s} \approx 58$ (but not in the $\bar p_t'$ trend of Fig.~\ref{mptxx}, left), corresponding to $\nu \approx 2.3$ as in Fig.~\ref{nchaa} (right). 
The main uncertainty in this figure occurs for peripheral collisions. The dash-dotted curve in the right panel describes the data trend below the ST resulting if the bold dotted trend for $x(n_s)$ in Fig.~\ref{nchaa} (right) were employed, which would seem to be consistent with the assignment $\bar \rho_{sNN}(n_s) \approx \bar \rho_{sNSD}$ required by  Fig.~\ref{nchaa} (left). However, that trend is inconsistent with the \pp\ $\bar p_{th}(n_s)$ trend in Fig.~\ref{alice5} (right) denoted by the bold dotted curves in this figure that might be expected for isolated \pn\ collisions in very peripheral \pa. The TCM is then consistent only if the relation $x(n_s) \approx \alpha \bar \rho_{sNN}(n_s)$ is relaxed in the \aa\ environment. As expected, jet formation is altered above the ST (``jet quenching''), but dijet production in \nn\ may already be strongly affected below the ST within peripheral \aa\ collisions.

\subsection{Uncertainties for A-B centrality determination}

The evolution of dijet production (both production rate and jet characteristics)  from ``in-vacuum'' \pp\ to peripheral \aa\ collisions is not well defined: Jet characteristics in \pp\ collisions are already strongly modified from truly in-vacuum (e.g.\ $q$-$\bar q$) jets from \ee\ collisions~\cite{eeprd,fragevo}, popular measures of jet characteristics such as $R_{AA}$ provide incomplete and possibly misleading information~\cite{mbdijets}, and determination of \aa\ centrality is least accurate for peripheral collisions~\cite{powerlaw} with no correspondence for (noneikonal) \pp\ collisions.

In principle, centrality (i.e.\ impact parameter $b$) for any composite collision system A-B might be derived from a Glauber Monte Carlo (MC) model simulation based on the eikonal approximation: partons within projectile nucleons or nucleons within nuclei travel along {\em resolved} (in the transverse plane) straight-line trajectories through the collision partner. However, yield, spectrum and correlation data indicate that the eikonal approximation does not apply to \pp\ collisions based on~\cite{ppquad} (a) the $\bar \rho_h\propto \bar \rho_s^2$ trend for MB dijet production and (b) no evidence for \pp\ eccentricity variation from a quadrupole correlation component. The concept of centrality or impact parameter is then not relevant to \pp\ collisions. Eventwise fluctuations in charge multiplicity and dijet production for \pp\ collisions may result from varying penetration depth of virtual splitting cascades on momentum fraction $x$ within projectile nucleons.

A Glauber-model determination of \aa\ centrality relies on correspondence between fractional cross section $\sigma / \sigma_0$ vs $N_{part}$ determined from the MC and  $\sigma / \sigma_0$ vs \nch\  inferred from a MB distribution on \nch. One directly observes $dP/dn_{ch}$, a probability (normalized event frequency) distribution on \nch. Any map from $dP/dn_{ch}$ to $d\sigma / dn_{ch}$ is model dependent, for instance based on the assumption that all \nn\ collisions are the same, or at least that all nucleon projectiles are the same. Centrality may be estimated e.g.\ by ``[a]ssuming that the average...multiplicity [$n_x$] is proportional to the number of participants [$N_{part}$] in an individual p–A collision...''~\cite{aliceppbprod}. 

However, as observed for \pp\ collisions eventwise nucleon structure in \nn\ collisions may fluctuate strongly, and any imposed ``centrality'' condition may then bias the number of gluon participants in \nn\ collisions.  The assumed correspondence between $n_x$ and $N_{part}$ in the quoted statement depends on how one defines ``participant'' and $n_x$. The statement may be correct if ``participants'' are low-$x$ gluons and $n_x$ is $n_s$, but is probably incorrect if ``participant'' means participant nucleon and $n_x$ is \nch. The fundamental issue is the nature of factorization in Eq.~(\ref{pafactor}). The \pa\ \mmpt\ data in this study indicate that the correspondence between measured probability (frequency) distribution $dP/dn_{ch}$ and A-B geometry in the form $(1/\sigma_0)d\sigma / dn_{ch}$ is quite uncertain, especially for small (\pp) and/or asymmetric (\pa) systems.

\subsection{Overall accuracy of the $\bf \bar p_t$ TCM}

This study demonstrates that within a TCM context precise high-statistics \mmpt\ data are easily and accurately described by simple functional forms interpretable in terms of standard QCD processes. Multiplicity dependence of \pp\ spectra over the largest possible interval is essential to achieve an accurate TCM.  A detailed picture emerges of the transition from noneikonal \nn\ to \pa\ and peripheral \aa\ to Glauber-dominated \aa. 

Good accuracy is achievable if a model is well matched in its degrees of freedom to the information carried by data. Key elements of the TCM were initially developed by model-independent {\em inductive} analysis of spectrum data wherein the simplicity of spectrum \nch\ dependence (only a few degrees of freedom) was discovered. Interpretation of model elements led to the realization that adopting soft component $n_s$ or $\bar \rho_s$ as the independent model parameter ensures a coherent description of \pp, \pa\ and \aa\ systems with low-$x$ gluons as the common basis, in isolated \nn\ collisions and in A-B collisions.

The accuracy and simplicity of the TCM as presented here is contrasted with certain assumptions regarding A-B collision geometry (e.g.\ applying a Glauber model to \pa\ data~\cite{aliceppbprod}), assumption of {\em total} \nch\ as the basic parameter~\cite{starnch} or introduction of complex, parametrized Monte Carlos based on {\em a priori} assumptions~\cite{pythia,mpitheory}.


\section{Discussion} \label{disc}

As noted in Ref.~\cite{alicempt} there are striking differences between \mmpt\ vs \nch\ trends in \pp\ and \aa\ collisions. \mmpt\ increases strongly with \nch\ for \pp\ collisions but much less rapidly for \aa\ collisions. The \pa\ trend is intermediate.  The \mmpt\ increase in \aa\ collisions is conventionally attributed to radial flow: ``In central Au+Au collisions the flattening of the spectra [and hence increase in \mmpt] is likely dominated by collective transverse radial flow, developed due to the large pressure buildup in the early stage of heavy-ion collisions''~\cite{radialflow}. That interpretation has been extended recently to \pa\ data to conclude that radial flow may be {\em larger} in smaller systems~\cite{dusling}.
In contrast, the TCM as applied to any A-B collision system provides a self-consistent description of two main hadron production mechanisms inferred from data: soft (projectile-nucleon dissociation) and hard (large-angle parton scattering to jets). The \mmpt\ soft component is universal -- fixed at $\bar p_{ts}\approx 0.4$ GeV/c and consistent with $h$-A fixed-target results~\cite{witpa,bialas}. Hard component $\bar p_{th}(n_s,\sqrt{s})$ corresponds to measured jet characteristics, with a noneikonal trend in \pp\ and a Glauber trend plus jet modification in \aa. A flow component is not required for the TCM data description. This study addresses apparent contradictions.

\subsection{Large $\bf \bar p_t$ values indicate MB dijets} \label{pbpbdetails}

Uncorrected $\bar p_t' \equiv \bar P_t' / n_{ch}'$ is an intensive ratio motivated in part by hopes to estimate the ``temperature'' of a thermalized QGP (e.g.\ measurement of eventwise mean-\pt\ fluctuations might test thermal equilibrium~\cite{na49ptfluct}). The intensive $\bar p_t'$ ratio tends to mix characteristics of two extensive variables $\bar P_t$ and $n_{ch}$ and includes the effects of a low-\pt\ acceptance cutoff (near 0.15 GeV/c for the data used in this study).  In contrast, extensive and corrected quantity $\bar P_t$ is more simply modeled (in terms of TCM soft component $\bar \rho_s$ as a basic parameter) and more interpretable, as demonstrated in this study. 

Given the introductory text for this section what does strong \mmpt\ variation and its relative extent in \pp, \pa\ and \aa\ systems imply? Is  \mmpt\ variation due to radial flow or MB dijets or some combination of other mechanisms? The basic issue is how longitudinal momentum in the initial state is transported to transverse phase space in the final state and with what efficiency: parton scattering to dijets vs strong multiple {\em re}scattering leading to flows. 

Jet production is a signature manifestation of QCD in high-energy nuclear collisions~\cite{bbk,fff}, whereas flows  (as conventionally described) rely on copious (parton and/or hadron) rescattering and high matter/energy densities to produce the large gradients required to drive flow phenomenon~\cite{hydro}. Thus, one should assume that jet production {is} the dominant transport mechanism until proven otherwise. That choice might resolve the counterintuitive conclusion of larger flows in smaller systems.

However, evidence from spectrum and correlation data provides a stronger argument in favor of jets. Jet-related angular correlations and the spectrum hard component from \pp\ collisions clearly manifest a $\bar \rho_h \propto \bar \rho_s^2$ multiplicity dependence followed precisely over a large multiplicity interval. In a jet context that trend can be interpreted as noneikonal response to a low-$x$ gluon flux represented by $\bar \rho_s$. The eikonal response would be $\bar \rho_h \propto \bar \rho_s^{4/3}$ as in the Glauber model~\cite{powerlaw}. Parameter $x(n_s) \equiv \bar \rho_h / \bar \rho_s$ measuring ``jettiness'' is then largest for noneikonal \pp\ collisions and smallest for eikonal \aa\ collisions with \pa\ collisions intermediate. The \mmpt\ data for several A-B systems actually inform us about centrality issues and dijet production as opposed to flow mechanisms, thus resolving the apparent paradox introduced by a flow interpretation.

For \aa\ collisions the relatively lower ``jettiness'' may make the final state appear softer, more ``thermal'' in accord with QGP expectations. But the absolute number of jets resolved in the final state can be much larger than in \pp\ or \pa. Both TCM parameters $x(n_s)$ and $\bar p_{thNN}(n_s)$ are expected to evolve with \aa\ centrality per jet modification, and those details for 2.76 TeV \pbpb\ collisions (one goal of the present study) are reported in Sec.~\ref{pbpb}.

Jet structure changes in \aa\ quantitatively in two ways: (a) Fragmentation functions are softened in a manner still described by the DGLAP equations~\cite{borg}. Only the gluon splitting function coefficient is increased, by 10\%~\cite{fragevo}. (b) The same-side 2D jet peak in angular correlations is modified by elongation on $\eta$ (but {\em not} on $\phi$) and possible development of long tails for higher parton energies~\cite{anomalous}. Those modifications are consistent with \mmpt\ TCM trends reported in Sec.~\ref{pbpbtrends} but are not consistent with expectations for jet ``quenching'' in a dense medium.

\subsection{Comparisons with Monte Carlo models}

In Ref.~\cite{alicempt} several Monte Carlo models are compared to \mmpt\ vs \nch\ data for three A-B systems: ``
For pp collisions, this [the strong increase in \mmpt\ with \nch] could be attributed, within a model of hadronizing strings, to multiple-parton interactions [MPIs] and to a final-state color reconnection [CR] mechanism. The data in p-Pb and Pb-Pb collisions cannot be described by an incoherent superposition of nucleon-nucleon collisions and pose a challenge to most [all] of the event generators.'' Three MCs are closely related: PYTHIA~\cite{pythia} is a popular \pp\ model emphasizing MPIs, with default and CR options.  HIJING~\cite{hijing} is default PYTHIA coupled to the Glauber model of \aa\ collision geometry. AMPT~\cite{ampt} is HIJING with string-melting and final-state rescattering options. Some relevant specifics of PYTHIA are addressed below.

The historical development of PYTHIA from the mid eighties has been reviewed recently~\cite{mpitheory}. Several critical assumptions were based on experimental observations during that period: (a) Almost all \pp\ events include at least one dijet -- MPIs represent the bulk of the nondiffractive cross section. That assumption rejects a minijet~\cite{ua1} cutoff scale near 5 GeV~\cite{ua1}: MPIs extend much lower than 5 GeV in order to give enough ``activity.'' (b) Essentially all hadrons near midrapidity come from jets, characterized by $n_{ch} \propto \bar n_{MPI}$, with $\bar n_{MPI}(p_{t,min})$ = $\sigma_{int}(p_{t,min}) / \sigma_{NSD}$ as the mean number of MPIs and $\sigma_{int}(p_{t,min})$ obtained by integrating the 2-2 jet spectrum $d\sigma_{int}(p_t) / d p_t^2$ down to some $p_{t,min} = O$(1-2) GeV/c. 

In order to achieve agreement with various forms of \pp\ data further assumptions are required including impact-parameter dependence and color reconnection among MPIs within color-singlet hadrons. (c) It is assumed that $\bar n_{MPI}(b) \propto \tilde O(b)$ where $\tilde O(b)$ is equivalent to thickness function $T_{AB}(b)$ for \aa\ collisions.  $\tilde O(b)$ is by analogy a measure of parton-parton binary encounters in a \pp\ collision (via eikonal approximation). (d) MPIs can be {\em merged in pairs} (CR) as determined by a free parameter, resulting in fewer jet fragments per dijet and increased \mmpt. In effect, more than one scattered parton may be associated with a string fragmenting to a single jet. 

The PYTHIA MC (essentially a {\em one}-component model) and the \pp\ TCM are effectively in direct opposition.  Whereas the TCM is based on observations that most \pp\ events are ``soft,'' with no dijet present, and $p_{t,min} \approx 3$ GeV/c~\cite{ppprd,fragevo,ppquad}, assumption (a) based on conjectured  $p_{t,min} \approx 1$-2 GeV/c states  that almost all \pp\ events contain at least one dijet. The two \pt\ limits represent a factor ten or more difference in jet production given $d\sigma_{int}(p_t) / d p_t^2 \sim 1/p_t^7$ near 3 GeV/c consistent with the spectrum trend in Fig.~\ref{pp1} (right)~\cite{jetspec2}. Whereas a {\em nonjet} soft component dominates hadron production in the TCM (consistent with $h$-A fixed-target observations~\cite{witpa,bialas}), assumption (b) within PYTHIA states that jet-related MPIs dominate hadron production. 

Whereas the TCM is based on an {\em observed} noneikonal dijet production trend $\propto \bar \rho_s^2$~\cite{ppprd,ppquad}, assumption (c) states that a conjectured eikonal dependence $\propto \tilde O(b)~ (\sim \bar \rho_s^{4/3})$ is included in PYTHIA. The latter leads to the trend $\bar p_t = \bar P_t / n_{ch} \propto n_{ch}^{1/3}$ as seen in Fig.~3 of Ref.~\cite{alicempt} for default PYTHIA, while the TCM results in  $\bar P_t / n_{s} \propto \bar \rho_s$ that describes data accurately with no additional assumptions as in Fig.~\ref{alice5a} of the present study. Whereas the TCM implicitly assumes independent dijet production, assumption (d) is based on strong MPI couplings via a CR mechanism. According to Ref.~\cite{mpitheory} if $n_{ch} \propto \bar n_{MPI}$ and jet-related $\bar P_t \propto \bar n_{MPI}$ then $\bar p_t = \bar P_t / n_{ch}$ should be independent of \nch, strongly disagreeing with \pp\ \mmpt\ data. Thus, some CR mechanism is required within PYTHIA.

The HIJING and AMPT Monte Carlos based on PYTHIA strongly disagree with \ppb\ and \pbpb\ data, in part because of PYTHIA structure and in part because jet modifications in \pbpb\ are not described correctly or at all. In contrast, the TCM provides an accurate description of \mmpt\ data in several A-B systems and is consistent with independent measurements of isolated jets and fixed-target $h$-A collisions dominated by soft physics.

\section{Summary} \label{summ}

This study addresses  an apparent contradiction in the interpretation of variation with charge multiplicity \nch\ of ensemble-mean transverse momentum \pt\ denoted by \mmpt. The conventional interpretation within the context of nucleus-nucleus (\aa) collisions is that increase of \mmpt\ with \nch\ signals radial flow due to gradients in a matter/energy density. But according to available \mmpt\ data the smaller the collision system the larger the radial flow (and hence gradients), and that conclusion contradicts the expectation that larger systems (i.e.\ heavy ion collisions) should develop greater densities and gradients and possibly form a locally-equilibrated quark-gluon plasma.

Alternatively, dijet production is the signature manifestation of QCD in high-energy nuclear collisions. Jet properties and production rates for elementary \pp\ collisions have been measured extensively. The systematics of minimum-bias (MB) jet contributions to \pp\ \pt\ spectra and two-particle correlations are now quantitatively understood and represented by a two-component (soft + hard) model (TCM) of hadron production near midrapidity. In this study the \pt\ spectrum TCM for \pp\ collisions is adapted to describe \mmpt\ data for \pp\ collisions and extended to describe \mmpt\ for a general A-B collision system.


The TCM for total $P_t$ production (integrated within some angular acceptance)  is based on measurements of dijet production, parton fragmentation and jet spectra. It is assumed that of two contributions to $\bar P_t = \bar P_{ts} + \bar P_{th}$ the soft component is a universal feature of high energy collisions corresponding to longitudinal dissociation of participant nucleons. The complementary hard component is entirely due to MB jet production in \nn\ collisions. An equivalent model is applied to charged-hadron production in the form $n_{ch} = n_s + n_h$, with $\bar p_t \equiv \bar P_t / n_{ch}$. The \pp\ \mmpt\ TCM incorporates the observation that dijet production in \pp\ collisions follows a {\em noneikonal} trend on \nch: production in \pp\ collisions increases approximately quadratically with \nch\ rather than $n_{ch}^{4/3}$ as expected for the eikonal Glauber model that describes \aa\ collisions.


The \pp\ \mmpt\ TCM framework for various collision energies is adapted in turn to describe data from \ppb\ and \pbpb\ collisions. The common parameter is soft integrated charge $n_s$ or charge angular density $\bar \rho_s$ interpreted to represent participant low-$x$ gluons. The derived TCM parameters are then $x(n_s)$, the ratio of hard to soft multiplicity in an average nucleon-nucleon (\nn) collision, $\nu(n_s)$, the mean number of \nn\ binary collisions per participant-nucleon pair  and $\bar p_{thNN}(n_s)$, the \mmpt\ hard component averaged over \nn\ binary collisions.

For \pp\ collisions $\nu = 1$ by definition, and $x(n_s) \approx \alpha(\sqrt{s}) \bar \rho_s$ reflects the noneikonal dijet production trend with $\alpha(\sqrt{s})$ defined accurately by a previous \pp\ \pt\ spectrum study. With those elements of the \pp\ TCM defined {\em a priori} the $\bar p_{thNN}(n_s,\sqrt{s})$ trends can be inferred from available \mmpt\ data and are found to be quantitatively consistent with the \pt\ spectrum hard component previously obtained from \pp\ data, which is in turn quantitatively consistent with measured properties of isolated jets. The noneikonal trend suggests that centrality is not relevant for \pp\ collisions, consistent with recent correlation data.

For \ppb\ collisions \mmpt\ data indicate that the noneikonal trend  $x(n_s) \approx \alpha(\sqrt{s}) \bar \rho_s$ persists up to transition point $\bar \rho_{s0}$ beyond which the linear trend continues but with 10-fold reduced slope to accommodate \mmpt\ data. Once $x(n_s)$ is defined the \pa\ Glauber number of participant pairs $N_{part}/2$, binary collisions $N_{bin}$ and binary collisions per participant pair $\nu \equiv 2N_{bin} / N_{part}$ follow by definition. The \ppb\ TCM transitions smoothly from a noneikonal \pp\ trend at lower multiplicities to a GLS model of \pa\ collisions, with no modification of jet fragmentation.

The \pbpb\ \mmpt\ data have several implications. Hard-component $\bar P_{th}$ production generally follows binary-collision scaling ($\propto N_{bin}$) according to the eikonal approximation as expected for dijet production in \aa\ collisions. However, above a sharp transition previously observed for \auau\ jet angular correlations parameter $x(n_s)$ increases rapidly and substantially (3-fold increase) to a saturation value, but hard component $\bar p_{thNN}(n_s,\sqrt{s})$ decreases by about 25\% also to a saturation value. Those trends are consistent with measured changes in jet properties (e.g.\ softened fragmentation functions) associated with ``jet quenching'' in more-central \aa\ collisions


To conclude, whereas attribution of \mmpt\ variation with \nch\ to radial flow leads to paradoxical conclusions within a flow/hydro context the TCM for \mmpt\ based on dijet production as the source of the \mmpt\ hard component accurately and consistently describes \mmpt\ data from several collision systems over a large range of collision energies and charge multiplicities, with only a few simple parameters. Low-$x$ gluons represented by $n_s$ or $\bar \rho_s$ provide a common basis for the TCM. The progression from fast to slow increase of \mmpt\ with \nch\  from \pp\ to \pa\ to \aa\ is found to result from collision geometry: noneikonal for \pp\ transitioning to full eikonal for \aa, with \pa\ as an intermediate case. The \mmpt\ hard component $\bar p_{tNN}(n_s,\sqrt{s_{NN}})$ corresponds quantitatively to measured \pt\ spectrum hard components which in turn correspond quantitatively to measured properties of isolated jets. Given the accuracy and simplicity of the \mmpt\ TCM and its direct connection with jet physics there seems to be no need to invoke a hydro mechanism for \mmpt\ variation in any A-B system.

\begin{appendix}

\section{TCM parametrizations} \label{equations}

Section~\ref{nchvar} describes adaptation of inferred \nch\ dependence of the TCM for \pp\ \pt\ spectrum hard components for 200 GeV and 13 TeV presented in Ref.~\cite{alicetomspec} to describe \nch\ dependence of $\bar p_{th}(n_s,\sqrt{s})$ data inferred in the present study. In this appendix details of the TCM parametrization for 200 GeV and 7 TeV are presented.

As noted in Sec.~\ref{nchvar}  the TCM spectrum hard component on transverse rapidity \yt\ is modeled by a Gaussian with exponential tail. The hard-component model shape is determined by separate Gaussian widths $\sigma_{y_t+}$ and $\sigma_{y_t-}$ above and below the peak mode and exponential parameter $q$, all varying with control parameter $\bar \rho_s = n_s / \Delta \eta$ to accommodate spectrum data as in Ref.~\cite{alicetomspec}. Figure~\ref{newparams1} summarizes parameter trends determined in that study. The same data points appear in Fig.~\ref{newparams2} below.

\subsection{200 GeV {\em p-p} parameters} \label{200gevparams}

\begin{figure}[h]
  \includegraphics[height=1.66in]{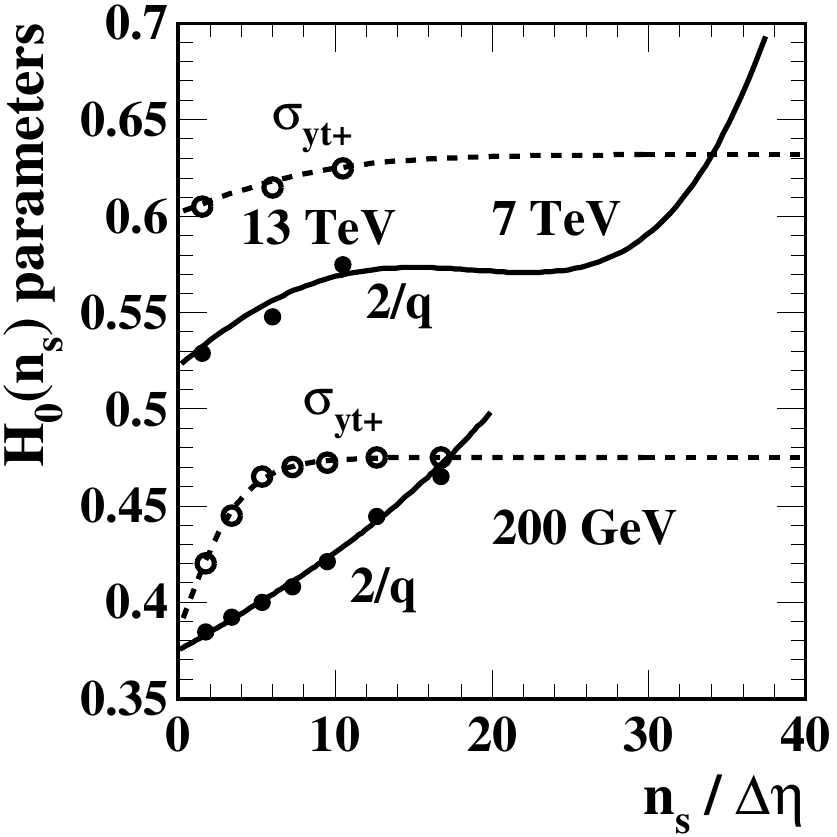}
   \includegraphics[height=1.63in]{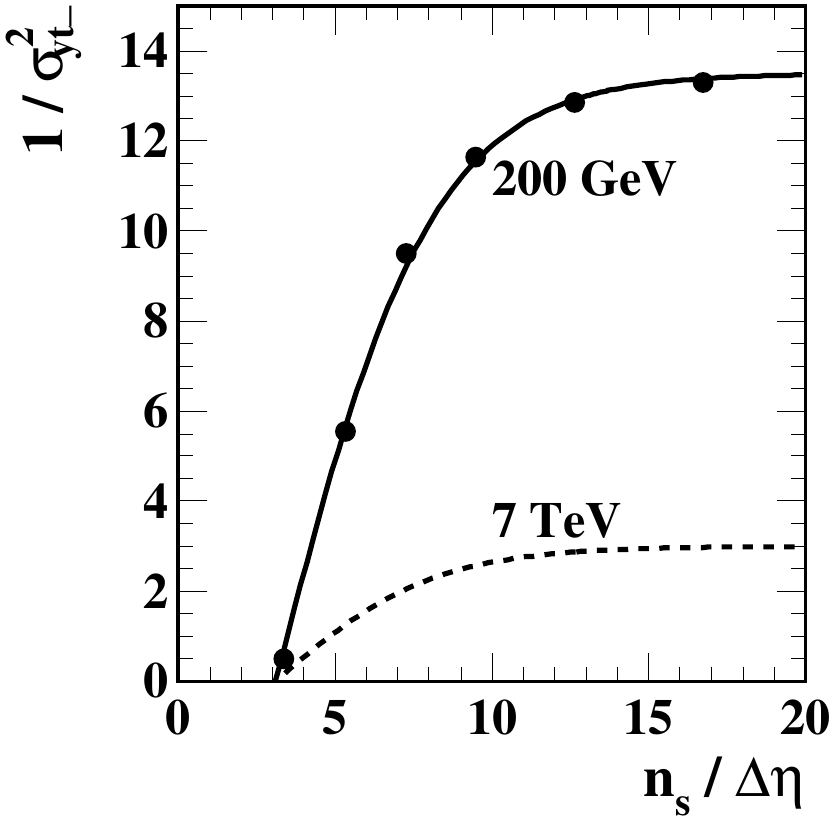}
\caption{\label{newparams2}
Left: TCM hard-component model parameters $\sigma_{y_t +}$ and $2/q$ (points) that accommodate spectrum data above the peak mode, inferred from \pt\ spectrum \nch\ dependence for 200 GeV and 13 TeV in Ref.~\cite{alicetomspec}. 
Right:  Model parameter $\sigma_{y_t-}$ (points) for multiplicity classes $n = 2$-7 of 200 GeV \pp\ collisions that accommodates spectrum data below the peak mode. The curves are defined in Eqs.~(\ref{200geveq}) and (\ref{7teveq}).
 }  
\end{figure}

The 200 GeV curves in Fig.~\ref{newparams2} are defined by
\bea \label{200geveq}
\sigma_{y_t +} &=& 0.385+0.09 \tanh(\bar \rho_s / 4)
\\ \nonumber
2/q &=& 0.373 + 0.0054\bar \rho_s ~~~\text{old}
\\ \nonumber
 &=& 0.375 + 0.0047(\bar \rho_s + 0.0008\bar \rho_s^3) ~~~\text{new}
\\ \nonumber
1 / \sigma^2_{y_t -} &=& 13.5\tanh[(\rho_s-3.1)/5].
\eea
Except as noted the expressions are just as in Ref.~\cite{alicetomspec}. The chosen forms of the plotted variables facilitate simple algebraic expressions. The 200 GeV parametrization is tightly constrained by the spectrum data over the same $\bar \rho_s = n_s / \Delta \eta$ range as the \mmpt\ data used for this study. The slight modification of the $q(n_s)$ parametrization does not significantly change the agreement in Fig.~\ref{300b} (right).

\subsection{7 TeV {\em p-p} parameters} \label{7tevparams}

The 7 TeV curves in Fig.~\ref{newparams2} are defined by
\bea \label{7teveq}
\sigma_{y_t +} &=& 0.60 + 0.03 \tanh(\bar \rho_s / 10)
\\ \nonumber
2/q &=& 1.4[0.373+0.0036 \bar \rho_s]~~~\text{old}
\\ \nonumber
2/q &=& 1.4[0.373+0.005 (\bar \rho_s - 0.026 \bar \rho_s^2 
\\ \nonumber
&& -~ 0.0012 \bar \rho_s^3   +  0.000044 \bar \rho_s^4)]~~~\text{new}
\\ \nonumber
1 / \sigma_{y_t -}^2 &=& 3\tanh[(\bar \rho_s-3.1)/5]~~~\text{new}
\eea
The 7 TeV parametrization is only loosely constrained by spectrum data and over a much smaller $\bar \rho_s$ range than the \mmpt\ data used for this study.  Commenting on the trends in Figs.~\ref{300b} and \ref{300a} (right), for $\bar \rho_s < 15$ the $\bar p_{th}(n_s)$ trend is dominated by $\sigma_{y_t-}(n_s)$. For $\bar \rho_s > 15$ the trend is dominated by $\sigma_{y_t+}(n_s)$ and $q(n_s)$, but for the higher energy only $q(n_s)$ matters because the Gaussian-exponential transition is close to the mode and the width above the mode $\sigma_{y_t +}$ plays no significant role. It is interesting that the form of $\sigma_{y_t -}$ at two widely-separated collision energies corresponds on $\bar \rho_s$ without scaling, even though $\bar \rho_{sNSD}$ is 2.5 in one case and 6 in the other case. The amplitude ratio corresponds to a factor-2 increase in $\sigma_{y_t -}$ at 7 TeV.

\end{appendix}


\end{document}